\documentclass[reprint,amsmath,amssymb,aps,superscriptaddress]{revtex4-1}
\usepackage{graphicx}
\usepackage{dcolumn}
\usepackage{bm}
\usepackage{url}
\usepackage{lipsum}
\usepackage{hyperref}
\usepackage[mathlines]{lineno}
\usepackage{upgreek}
\usepackage{biolinum}

\newcommand{\pxp}{p_x{>}0}
\newcommand{\pxn}{p_x{<}0}
\newcommand{\pt}{p_{\rm T}}
\newcommand{\etap}{0{<}\eta{<}2}
\newcommand{\etan}{-2{<}\eta{<}0}

\begin{document}

\title{Left-right splitting of elliptic flow due to directed flow in
  heavy ion collisions} 
\author{Chao Zhang}
\affiliation{Department of Physics, East Carolina University, Greenville, NC 27858, USA}
\affiliation{ Institute of Particle Physics and Key Laboratory of
  Quark \& Lepton Physics (MOE), Central China Normal University, Wuhan
  430079, China} 
\author{Zi-Wei Lin}
\email{linz@ecu.edu}
\affiliation{Department of Physics, East Carolina University,
  Greenville, North Carolina 27858, USA}

\begin{abstract}
Recently the splitting of elliptic flow $v_2$ at finite rapidities
has been proposed as a result of the global vorticity in non-central
relativistic heavy ion collisions.
In this study, we find that this left-right
(i.e., on opposite sides of the impact parameter axis) splitting of
the elliptic flow at finite rapidities is a result of the non-zero
directed flow $v_1$, with the splitting magnitude $\approx
8v_1(1-3v_2)/(3\pi)$. We also use a multi-phase transport model, which
automatically includes the vorticity field and flow fluctuations, to 
confirm the $v_2$ splitting. 
In addition, we find that the analytical expectations for the $v_2$
splitting work for the raw $v_2$ and $v_1$ (i.e., before event
plane resolutions are applied) measured relative to either the first- or
second-order event plane. Since the $v_2$ splitting is mostly
driven by $v_1$, it vanishes at zero transverse momentum ($p_{\rm
  T}$), and its magnitude and sign may have non-trivial dependencies on
$p_{\rm T}$, centrality, collision energy, and hadron species. 
\end{abstract}

\maketitle

\section{Introduction}

Heavy ion collisions provide us a way to study the properties of the
created matter in relation to the theory of quantum
chromodynamics. At high-enough energies, the created matter should be
initially deconfined and be represented by parton degrees 
of matter, the quark-gluon plasma (QGP)
~\cite{Shuryak:1978ij}. Anisotropic flows in the final particle
momentum distribution are powerful tools to study heavy ion physics
~\cite{Voloshin:1994mz}. They include the directed flow $v_1$,
elliptic flow $v_2$, the more recently discovered triangular flow $v_3$
~\cite{Alver:2010gr}, and higher-order flows. 
Significant anisotropic flows have been observed in
non-central heavy ion collisions
~\cite{STAR:2000ekf,PHENIX:2003qra,ALICE:2010suc}, while they have also been
observed in small systems such as $p$+Au, $d$+Au, He+Au
~\cite{PHENIX:2018lia}, and $p$+Pb~\cite{CMS:2013jlh} collisions at high
energies.  It is well known that the initial geometry of the overlap
volume in a heavy ion collision has a fluctuating and complicated
event-by-event three-dimensional structure, and the initial spatial 
anisotropies are converted into anisotropic flows in momentum  
through particle interactions in transport models or the  pressure
gradient in hydrodynamical models. 
On the other hand, the origin of flow-like behavior in small
systems including $p+p$ collisions is still under debate, where 
initial-state correlations~\cite{Dusling:2017dqg,Mace:2018vwq},
the parton escape mechanism~\cite{He:2015hfa,Lin:2015ucn} 
or kinetic theory~\cite{Kurkela:2018ygx,Kurkela:2019kip}, and
hydrodynamics~\cite{Weller:2017tsr,Heinz:2019dbd} have been proposed
as the dominant mechanism.

Recently, the splitting of elliptic flow on opposite sides of the
impact parameter axis has been proposed as a new observable 
~\cite{Chen:2021wiv}, where it is argued to be the result of the
global vorticity in non-central relativistic heavy ion collisions. In
this study, we examine this $v_2$ splitting in A+A collisions at RHIC
and LHC energies.

\section{Method}

After a brief analysis of the left-right $v_2$ splitting 
that relates it to $v_1$, we shall use a multi-phase transport
(AMPT) model~\cite{Lin:2004en,Zhang:2021vvp} to test the relationship. 
The AMPT model is an event generator for relativistic heavy ion
collisions that contains four parts: the fluctuating initial
condition based on the HIJING model, the parton cascade
model ZPC for elastic scatterings, a spatial quark coalescence model
to describe the hadronization, and a hadron cascade.
The string melting version of the AMPT model
~\cite{Lin:2001zk,Lin:2004en} is applicable at high energies when the
QGP is expected to be formed. It has been shown to well describe the
bulk matter including the overall elliptic
flow~\cite{Lin:2001zk,Ma:2016fve} and the vorticity
field~\cite{Jiang:2016woz,Li:2017slc}  in heavy ion collisions. Its
fluctuating initial condition also creates spatial anisotropies that
lead to anisotropic flows of different orders~\cite{Alver:2010gr}. 
Recently we have updated the AMPT model with a new quark coalescence
model~\cite{He:2017tla}, modern parton distribution 
functions in nuclei~\cite{Zhang:2019utb},
and improved heavy quark productions~\cite{Zheng:2019alz}.
In addition, we have applied local nuclear scaling to two key parameters
in the AMPT initial condition, which enables the model to 
self-consistently describe the system size and centrality dependence of nuclear
collisions~\cite{Zhang:2021vvp}.
For this study, we employ the string melting version of the AMPT model
that contains the above improvements. 

In the AMPT model~\cite{Lin:2004en},
the $x$-axis is along the impact parameter ($\vec b$) in the
transverse plane of each event, 
where the centers of the projectile and target nuclei 
are at the transverse position $(b/2,0)$ and $(-b/2,0)$, respectively. 
The direction of the projectile momentum specifies the $z$-axis; 
as a result, the total angular momentum of the system is along the
$-y$ direction~\cite{Jiang:2016woz}. Note that this coordinate system
is the same as that in the recent study 
that first proposed the $v_2$ splitting~\cite{Chen:2021wiv}.
In Sec.\ref{RP-results}, we analyze the $v_2$ splitting and
calculate flows using the theoretical reaction plane for simplicity. 
Then in Sec.\ref{EP-results} we use the experimental event plane
method to calculate the flows and $v_2$ splitting.

\section{Results using the reaction plane}
\label{RP-results}

Let us write the normalized azimuthal distribution in momentum for
particles in a given phase space, e.g., at a given $\pt$ and rapidity
$y$ or pseudorapidity $\eta$, as
\begin{eqnarray}
 f(\phi)=\frac{1}{2\pi} \left ( 1+\sum_{n=1}^\infty \left [c_n \cos
  (n\phi)+s_n \sin (n\phi) \right ] \right ).  
\end{eqnarray}
The above gives the usual relations $\langle \cos \phi \rangle=c_1/2
\equiv v_1$ and $\langle \cos(2\phi) \rangle=c_2/2 \equiv v_2$ 
when we integrate over the full range of $\phi$ in the averaging over
particles and events.
On the other hand, if we integrate over $\phi \in (-\pi/2,\pi/2)$ for
the $\pxp$ part or over $\phi \in (\pi/2,3\pi/2)$ for the $\pxn$ part,
we obtain the following for $v_2$ (with terms up to $n=4$): 
\begin{equation}
\begin{split}
  v_2(\pxp) &\equiv \frac{\int_{-\pi/2}^{\pi/2}\cos(2\phi)
  f(\phi)d\phi}{\int_{-\pi/2}^{\pi/2}f(\phi)d\phi} 
  =\frac{v_2+\frac{4v_1}{3\pi}+\frac{6c_3}{5\pi}}
  {1+\frac{4v_1}{\pi}-\frac{2c_3}{3\pi}}, \\
  v_2(\pxn)
  &\equiv \frac{\int_{\pi/2}^{3\pi/2}\cos(2\phi)
  f(\phi)d\phi}{\int_{\pi/2}^{3\pi/2}f(\phi)d\phi} 
  =\frac{v_2-\frac{4v_1}{3\pi}-\frac{6c_3}{5\pi}}
  {1-\frac{4v_1}{\pi}+\frac{2c_3}{3\pi}}.
\end{split}
\label{v2s0}
\end{equation}
Note that the coefficient $c_3=2\langle \cos(3\phi) \rangle$ 
does not correspond to (twice) the usual triangular flow $v_3$
~\cite{Alver:2010gr}, which axes fluctuate mostly independently of the
$x$-axis (or the reaction plane). Instead, $c_3$ here is calculated
relative to the reaction plane and represents another type of triangular
flow at finite rapidities that correlates with the reaction plane, and
it is expected to be rapidity-odd. 
If $|c_3|\ll |v_1|$, we can neglect $c_3$ and get simpler relations
(up to second order in $v_1$ and/or $v_2$): 
\begin{equation}
\begin{split}
  v_2(\pxp) &\simeq \frac{v_2+\frac{4v_1}{3\pi}}{1+\frac{4v_1}{\pi}}
  \simeq v_2 - \frac{16v_1^2}{3\pi^2} +\frac{4v_1 }{3\pi}\left (1-3v_2
  \right  ), \\
  v_2(\pxn) &\simeq
  \frac{v_2-\frac{4v_1}{3\pi}}{1-\frac{4v_1}{\pi}}\simeq   v_2 -
  \frac{16v_1^2}{3\pi^2}-\frac{4v_1}{3\pi} \left (1-3v_2 \right ).
\end{split}
\label{v2s}
\end{equation}
Therefore, the left-right $v_2$ splitting at finite rapidities, 
given by $v_2(\pxp)-v_2(\pxn) \simeq 8v_1(1-3v_2)/(3\pi)$, 
comes directly from the finite directed flow $v_1$.

\begin{figure}[!htb]
\includegraphics[scale=0.45]{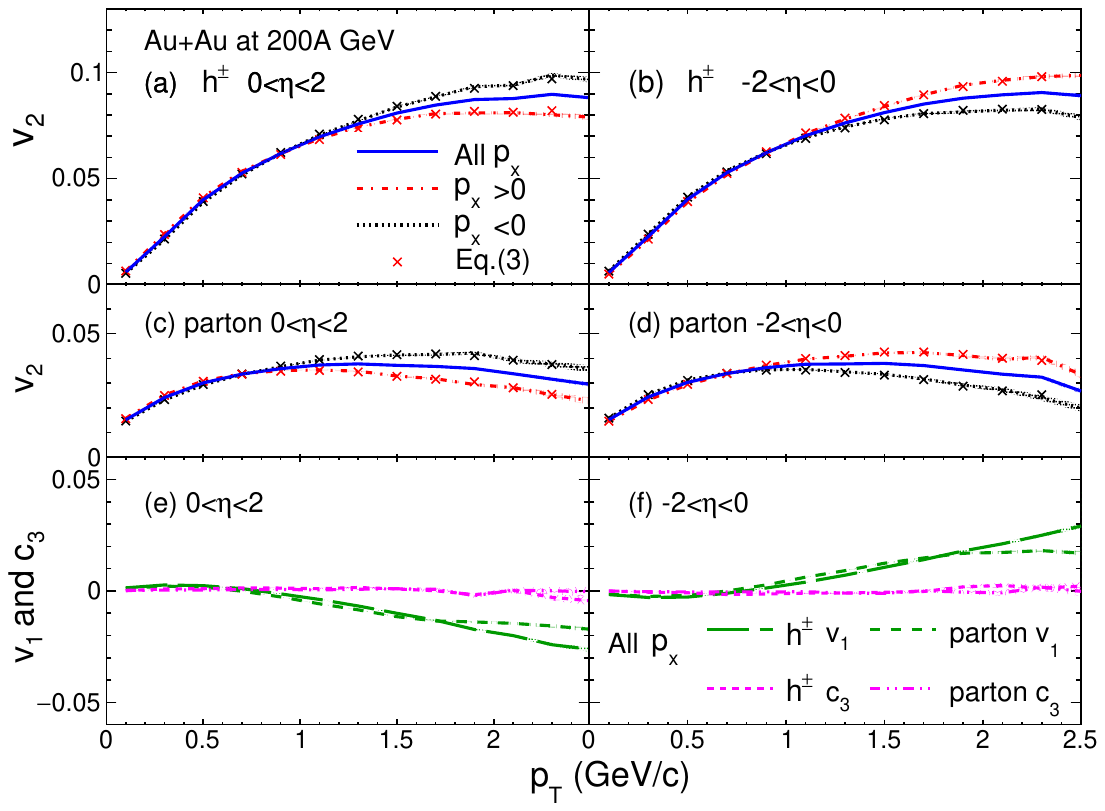}
\caption{Left panels show (a) $v_2$ of charged hadrons, (c) $v_2$ of partons,
(e) $v_1$ and $c_3$ within $\etap$ versus the transverse
momentum for minimum bias Au+Au collisions at $200A$ GeV
from the AMPT model. 
The overall $v_2$ (solid), $v_2(\pxp)$ (dot-dashed), and $v_2(\pxn)$
(dotted) are shown, while crosses represent the expectations of
Eq.\eqref{v2s}. The corresponding results for particles within
$\etan$ are shown in the right panels.}
\label{fig:1}
\end{figure}
 
We now use the AMPT model to test these relationships. 
Figure~\ref{fig:1} shows the results of $v_2(\pt)$ of charged 
hadrons and final state partons with $\pxp$, $\pxn$, and all $p_x$ for
minimum bias Au+Au collisions at $200A$ GeV. 
Results for the pseudorapidity ranges of $\etap$ and $\etan$ 
are shown in  the left and right panels, respectively. 
We see that the left-right splitting of elliptic flow exists for
both partons and hadrons in the final state. 
The $v_2$ splitting is small at low $\pt$ and then become obvious for
partons at $\pt > 1$ GeV/$c$ and  for hadrons at $\pt > 1.5$ GeV/$c$. 
In addition, we see that the $v_2$ splitting is antisymmetric versus
$\eta$, which is mostly due to the rapidity-odd $v_1$ as 
$c_3 \simeq 0$ in this figure.  
Note that we call this the left-right splitting of the elliptic flow
$v_2$, because particles with $\pxp$ and $\pxn$ are on the right and
left side of the impact parameter axis, respectively. 
This also helps to differentiate it from the $v_2$ splitting between
particles and antiparticles, which has been shown to be sensitive to
partonic and hadronic potentials at low energies or high net-baryon
chemical potential~\cite{Xu:2012gf, Xu:2013sta}.

We also show in Figs.~\ref{fig:1}(e) and (f) the overall $v_1(\pt)$ of
charged hadrons and partons at positive and negative pseudorapidities,
respectively. The expectations from Eq.\eqref{v2s} using the overall
(i.e., all $p_x$) $v_1(\pt)$ and $v_2(\pt)$ are shown by the cross
symbols in Figs.~\ref{fig:1}(a-d), which agree well with the 
directly calculated $v_2$ splitting curves. 
The good agreements with Eq.\eqref{v2s} in Fig.~\ref{fig:1} also
indicate $c_3(\pt) \simeq 0$ there, which we confirm in
Figs.~\ref{fig:1}(e) and (f) with direct calculations of $c_3(\pt)$
(magenta curves).

\begin{figure}[!htb]
\includegraphics[scale=0.43]{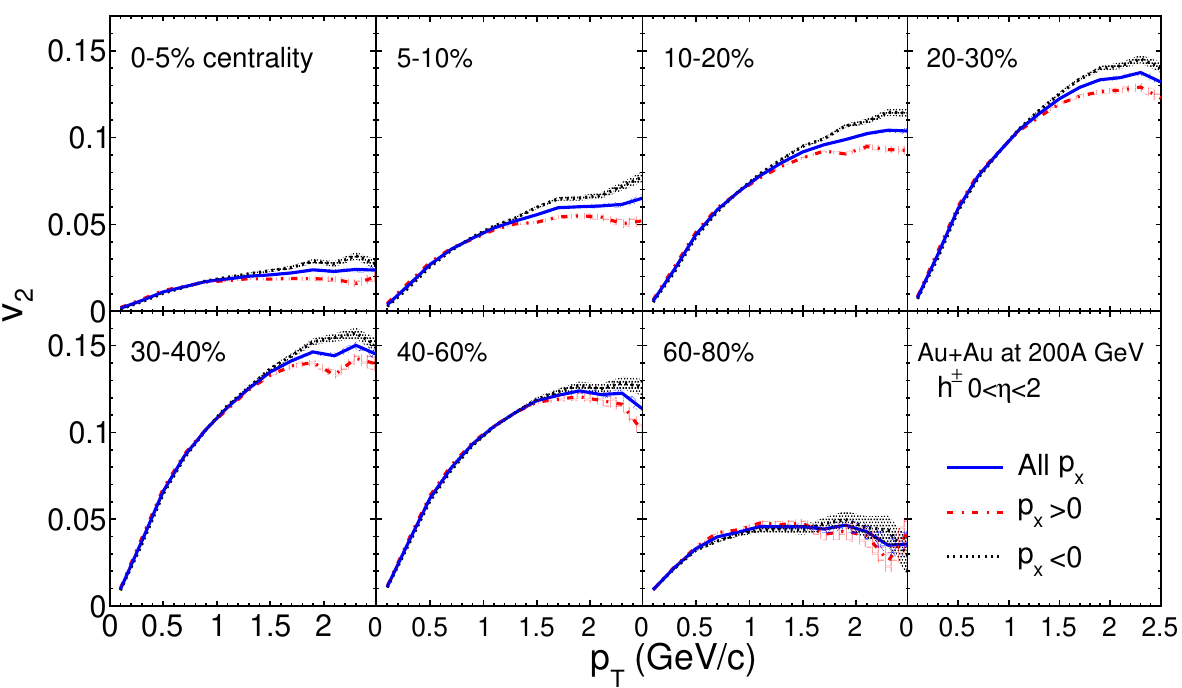}
\caption{The overall and split $v_2(\pt)$ of charged hadrons
  within $\etap$ for different centralities of Au+Au collisions
  at $200A$ GeV from the AMPT model.}
\label{fig:2}
\end{figure}

Next we investigate the possible centrality dependence of the $v_2$
splitting in Fig.~\ref{fig:2}, which shows the AMPT results of
$v_2(\pt)$ for charged hadrons within $\etap$ for different
centralities of Au+Au collisions at $200A$ GeV.
Here centrality is determined with the charged hadron 
multiplicity within mid-pseudorapidity. 
We observe the $v_2$ splitting at all centralities (except for very
peripheral collisions where the statistical errors are relatively
large). In addition, the order of the splitting at higher $\pt$, i.e., 
the sign of $v_2(\pxp)-v_2(\pxn)$, does not change with centrality
here. It is also interesting to see that the $v_2$ splitting exists
even in the most central (0-5\%) events.

\begin{figure}[!htb]
\includegraphics[scale=0.42]{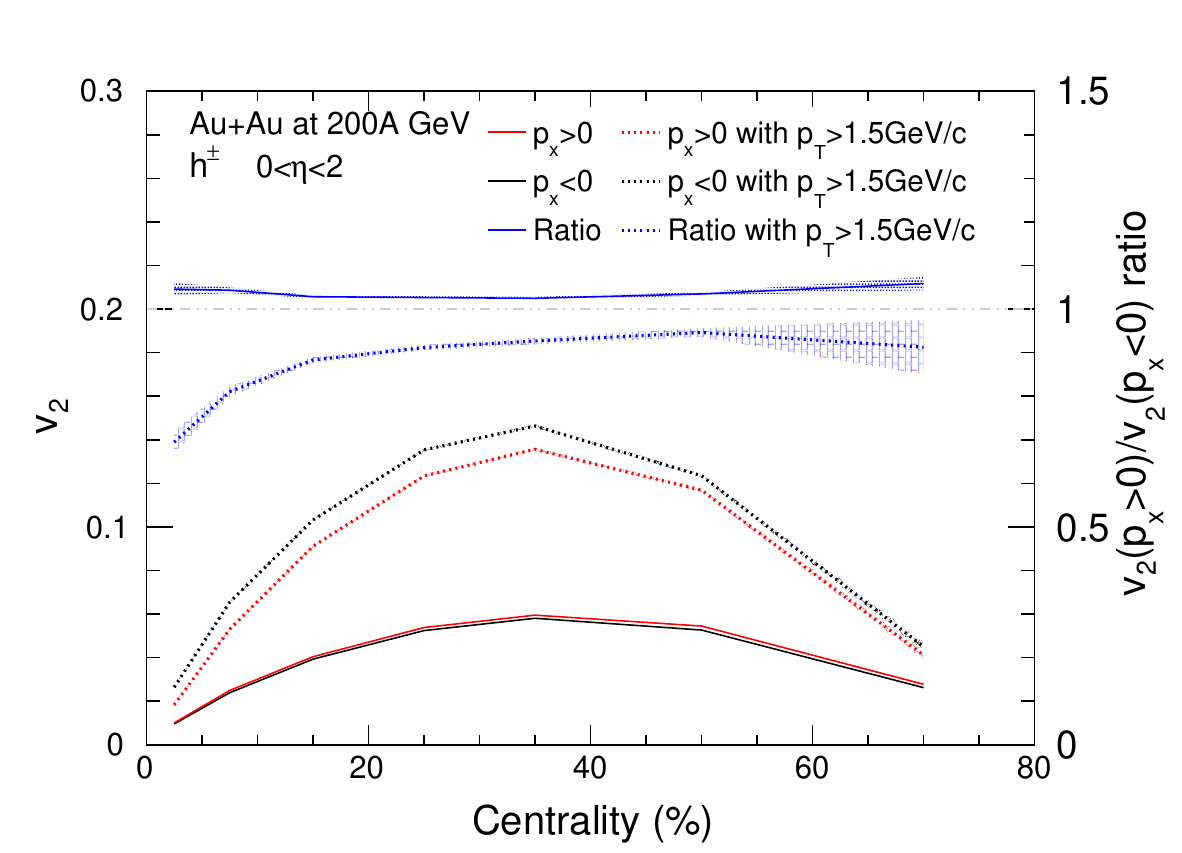}
\caption{Centrality dependence of $\pt$-integrated
  $v_2$ for charged hadrons with positive or negative
  $p_x$ and their ratio (blue curves). Solid and dotted $v_2$ curves  
  represent the results integrated over all $\pt$ and over $\pt>1.5$
  GeV/$c$, respectively.}
\label{fig:3}
\end{figure}

Figure~\ref{fig:3} shows the centrality dependence of the integrated 
$v_2$ of charged hadrons with positive or negative $p_x$ in Au+Au
collisions at $200A$ GeV from the AMPT model.
The $v_2$ integrated over $\pt>1.5$ GeV/$c$ shows a significant
left-right splitting, consistent with that shown in Fig.~\ref{fig:2}. 
On the other hand, the $v_2$ integrated over all $\pt$ shows a very
small splitting, while the order of splitting is also opposite to that of
the $v_2$ integrated over $\pt>1.5$ GeV/$c$.  This is because the
directed flow $v_1$ from the AMPT model changes sign at $\pt \sim 
0.7$ GeV/$c$, as shown in Fig.~\ref{fig:1}. 
We also show in Fig.~\ref{fig:3} the ratio $v_2(\pxp)/v_2(\pxn)$ to
represent the relative left-right $v_2$ splitting. 
While the ratio of the split $v_2$ integrated over all $\pt$ is rather flat
versus centrality, the ratio of the split $v_2$ integrated over $\pt>1.5$
GeV/$c$ shows a significant dependence on centrality, where
the relative splitting is the biggest for the most central
events. This is possible because, although both $v_2$ and $v_1$ 
usually approach zero for central collisions, the relative $v_2$
splitting mostly depends on $v_1/v_2$, which could be large.

\begin{figure}[!htb]
\includegraphics[scale=0.43]{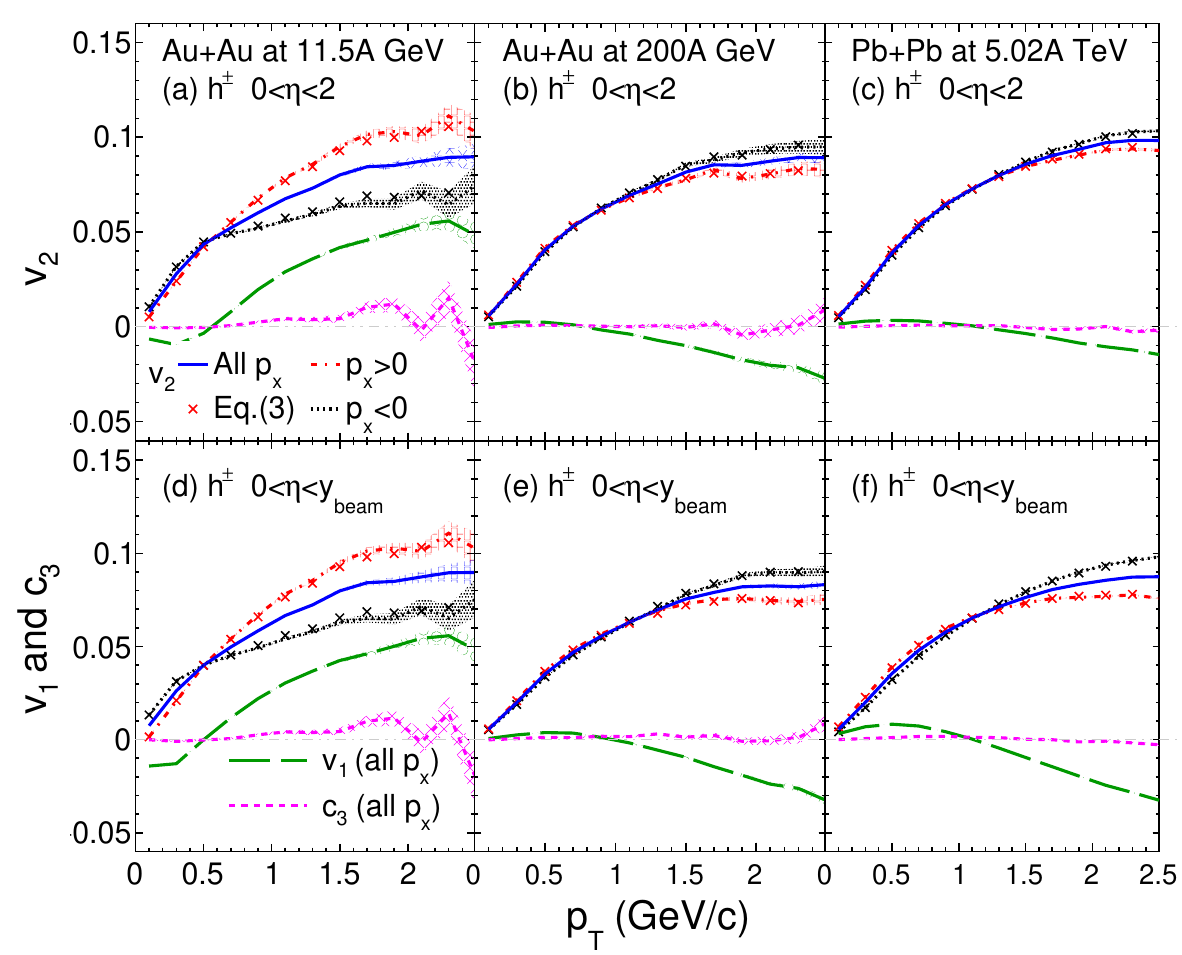}
\caption{The overall $v_2$ (solid), $v_2(\pxp)$ (dot-dashed), 
  $v_2(\pxn)$ (dotted), $v_1$ (long-dashed), and $c_3$
  (dashed) of charged hadrons within $\etap$  in  minimum bias (a)
  Au+Au collisions at $11.5A$, (b) $200A$ GeV and (c) Pb+Pb
  collisions at $5.02A$ TeV. The corresponding results within
  $0{<}\eta{<}y_{\rm beam}$ are shown in the lower panels, while
  crosses represent the expectations of Eq.\eqref{v2s}.}
\label{fig:4}
\end{figure}

In Fig.~\ref{fig:4} we examine the $v_2$ splitting at different
energies at RHIC and LHC, where the upper panels (a-c) show the AMPT
$v_2(\pt)$ results of charged hadrons within $\etap$ in minimum bias
Au+Au collisions at $11.5A$ and $200A$ GeV and minimum bias Pb+Pb
collisions at $5.02A$ TeV. 
The magnitude of the $v_2$ splitting decreases as the colliding energy
increases; this is expected because the directed flow $v_1$ is usually
smaller at higher energies.
We also see that the order of splitting can be different at different
energies as a result of the sign of $v_1(\pt)$. 
The expectations from Eq.\eqref{v2s}, shown as the cross symbols,
agree rather well with the split $v_2$ results at all three
energies, although there is a small discrepancy at $11.5A$ GeV. This
discrepancy is due to the $c_3$ terms in Eq.\eqref{v2s0}, because
$c_3$ is not very small at $11.5A$ GeV as shown in
Fig.~\ref{fig:4}(a). In cases where $|c_3| \ll |v_1|$ is not true,
Eq.\eqref{v2s} is not accurate enough while Eq.\eqref{v2s0} still is. 
In the lower panels of Fig~\ref{fig:4}, we show the split $v_2(\pt)$
results for charged hadrons within $0{<}\eta{<}y_{\rm beam}$, where
$y_{\rm beam}$ is the projectile rapidity in the center-of-mass frame
of the heavy ion collision.  
We see that the magnitudes of the $v_2$ splitting are similar at these
different energies, and the wider $\eta$ range increases the
magnitude of $v_2$ splitting at high energies due to the big $v_1$
magnitudes at large rapidities. This suggests that the left-right
$v_2$ splitting could be observed even at LHC energies if one looks at
large rapidities. 

\section{Results using the event plane}
\label{EP-results}

So far we have used the theoretical reaction plane (RP) for the flow
analysis. We now use the experimental event-plane
method to see whether the $v_2$ splitting can be observed
experimentally. Specifically, we use minimum bias Au+Au collisions at
$11.5A$ GeV as the example. 
The event plane angle $\Psi_n$ is calculated with the
event-by-event flow vector
$\vec{Q}_n$~\cite{Poskanzer:1998yz,Voloshin:2008dg} according to
\begin{equation}
\begin{split}
&X_n=\sum_i w_i\cos(n\phi_i),~Y_n=\sum_i w_i\sin(n\phi_i), \\	
&\Psi_n = {\rm arctan2}(Y_n,X_n)/n.
\end{split}
\end{equation}
In the above, $\phi_i$ is the azimuthal angle of the $i^{th}$
particle's momentum in the event, and the weight $w_i$ is taken as the
particle transverse momentum. The first-order event plane angle
$\Psi_1$ is reconstructed from two flow vectors: $\vec{Q}_{1,\rm
  west}$ from $3.5<\eta<5.09$ and $\vec{Q}_{1,\rm  east}$ from
$-5.09<\eta<-3.5$.   The first-order flow vector of the full event is
then reconstructed   as  $\vec{Q}_1=\vec{Q}_{1,\rm
  west}-\vec{Q}_{1,\rm east}$.  
Note that we have chosen a narrower $\eta$ range than 
the STAR event plane detector (EPD) system~\cite{Adams:2019fpo}
to get a better resolution of $\Psi_1$.
The second-order event plane angle $\Psi_2$ is reconstructed from the
flow vector of particles within $-2<\eta<2$.

\subsection{Relative to the first-order event plane $\Psi_1$}
\label{Psi1}

Flow coefficients relative to the first-order event plane are
calculated as 
\begin{equation}
v_n\{\Psi_1\}\equiv\frac{v_n^{\rm obs}}{R_n\{\Psi_1\}} = \frac{
  \langle\langle \cos [n(\phi-\Psi_1)] \rangle\rangle} 
{\langle \cos[n(\Psi_1-\Psi_{\rm r})] \rangle}. 
\end{equation}
In the above, $\Psi_{\rm r}$ represents the ``true'' reaction plane angle, 
and the numerator represents the observed (or raw) anisotropic flows which
are directly measured in the experiments,  where the double brackets 
represent the averaging over particles in each event and then over all
events.  The denominator represents the event plane resolution, where
the single bracket represents the averaging over all events.
We follow the two-subevent method~\cite{Poskanzer:1998yz} to calculate
the resolutions, where the two subevents are from $3.5<\eta<5.09$ and
$-5.09<\eta<-3.5$, respectively, for calculating $R_n\{\Psi_1\}$. Note
that we use a modified equation $R_{2k+1}^{\rm   sub}\{\Psi_1\}=\sqrt
{<\cos[(2k+1)(\Psi_1^a-\Psi_1^b-\pi)]>}$  instead of $\sqrt 
{<\cos[(2k+1)(\Psi_1^a-\Psi_1^b)]>}$~\cite{Poskanzer:1998yz}  to
calculate the subevent resolution for the odd-order anisotropies,  
because $\Psi_{1,\rm west}$ for one subevent is expected to be different
from $\Psi_{1,\rm east}$ for the other subevent by $\pi$ on average. 
We then determine the $\chi_n^{\rm sub}$ value from $R_n^{\rm sub}$. 
Using $\chi_n=\sqrt{2}\chi_n^{\rm sub}$ for the full
event~\cite{Poskanzer:1998yz}, we then obtain the full event
resolutions $R_n\{\Psi_1\}$.

\begin{figure}
\includegraphics[scale=0.45]{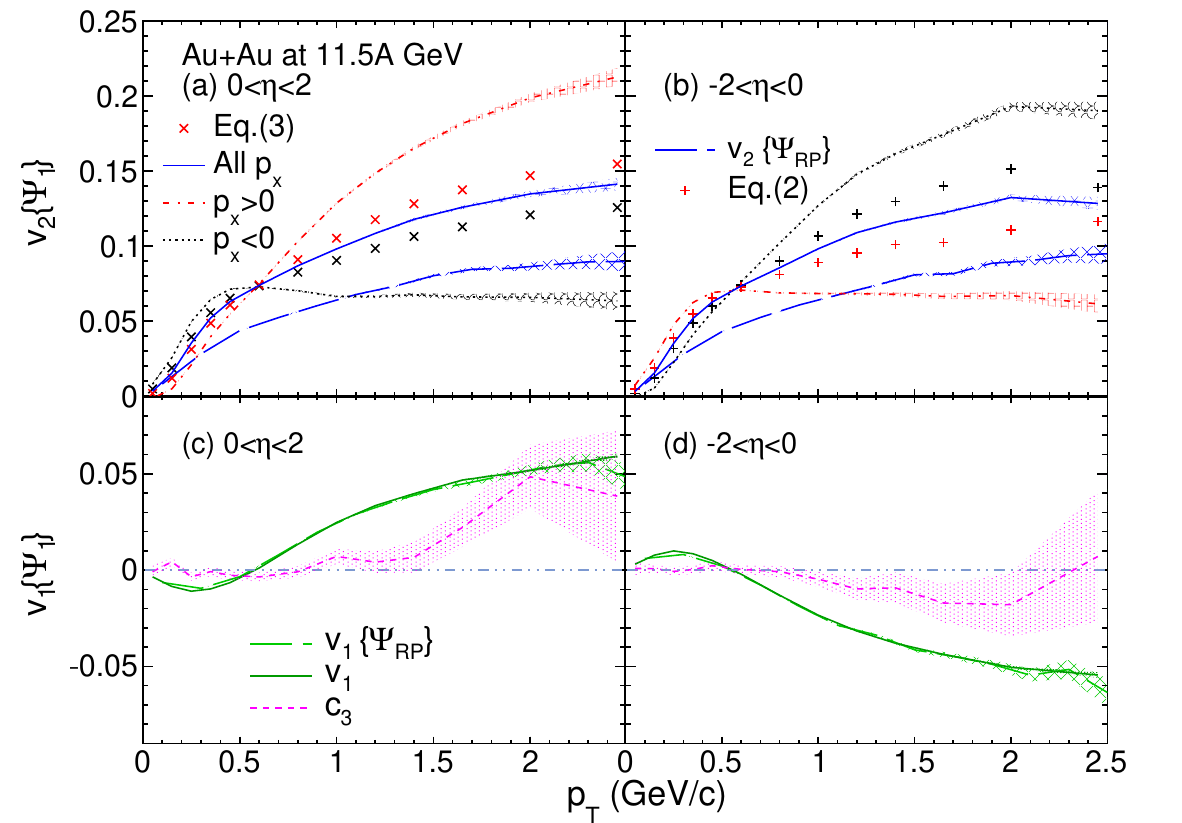}
\caption{(a) The overall and split $v_2$, and (c) $v_1$ and $c_3$ of
  charged hadrons relative to the first-order event plane angle
  $\Psi_1$ versus the transverse momentum   for $\etap$ in minimum
  bias Au+Au   collisions at $11.5A$   GeV from the AMPT
  model. Results for $\etan$   are shown in the right panels. Cross
  and plus symbols represent  expectations of Eq.\eqref{v2s} and
  Eq.\eqref{v2s0}, respectively,   and  long dashed curves represent
  $v_2$ (blue) and $v_1$   (green) relative to the theoretical
  reaction plane angle.} 
\label{fig:5}
\end{figure}

We show in Fig.~\ref{fig:5} the $v_2$ splitting results relative to
the first-order event plane after applying the resolution corrections 
with $R_n\{\Psi_1\}$. 
Note that here we treat the $\Psi_1$ angle as the $x$-axis, where any
hadron with $\cos(\phi-\Psi_1) > 0$ belongs to the $\pxp$ group.  The
upper two panels show the overall, left-side and right-side $v_2$ of
charged particles within $0<\eta<2$ and $-2<\eta<0$, respectively,
while the corresponding $v_1$ and $c_3$ results are 
shown in the lower panels. The cross symbols in Fig.~\ref{fig:5}(a)
are the expectations of the split $v_2$ by applying Eq.\eqref{v2s} to
the overall $v_2$ in panel (a) and overall $v_1$ in panel (c). 
It is clear that the expectations of Eq.\eqref{v2s} fail here. 
Since $c_3$ is not too small here, we also apply Eq.\eqref{v2s0} to
the overall $v_2$ in panel (b), and overall $v_1$ and overall $c_3$ in
panel (d) to obtain the plus symbols; they still fail to describe the
split $v_2$ curves in panel (b).  In addition, the long dashed curves
in Fig.~\ref{fig:5} are the $v_1\{\Psi_{\rm RP}\}$ and $v_2\{\Psi_{\rm 
  RP}\}$ relative to the theoretical reaction plane, where 
$v_1\{\Psi_{1}\}$ well reproduces $v_1\{\Psi_{\rm RP}\}$ while
$v_2\{\Psi_{1}\}$ are different from $v_2\{\Psi_{\rm RP}\}$.  

\begin{figure}
\includegraphics[scale=0.45]{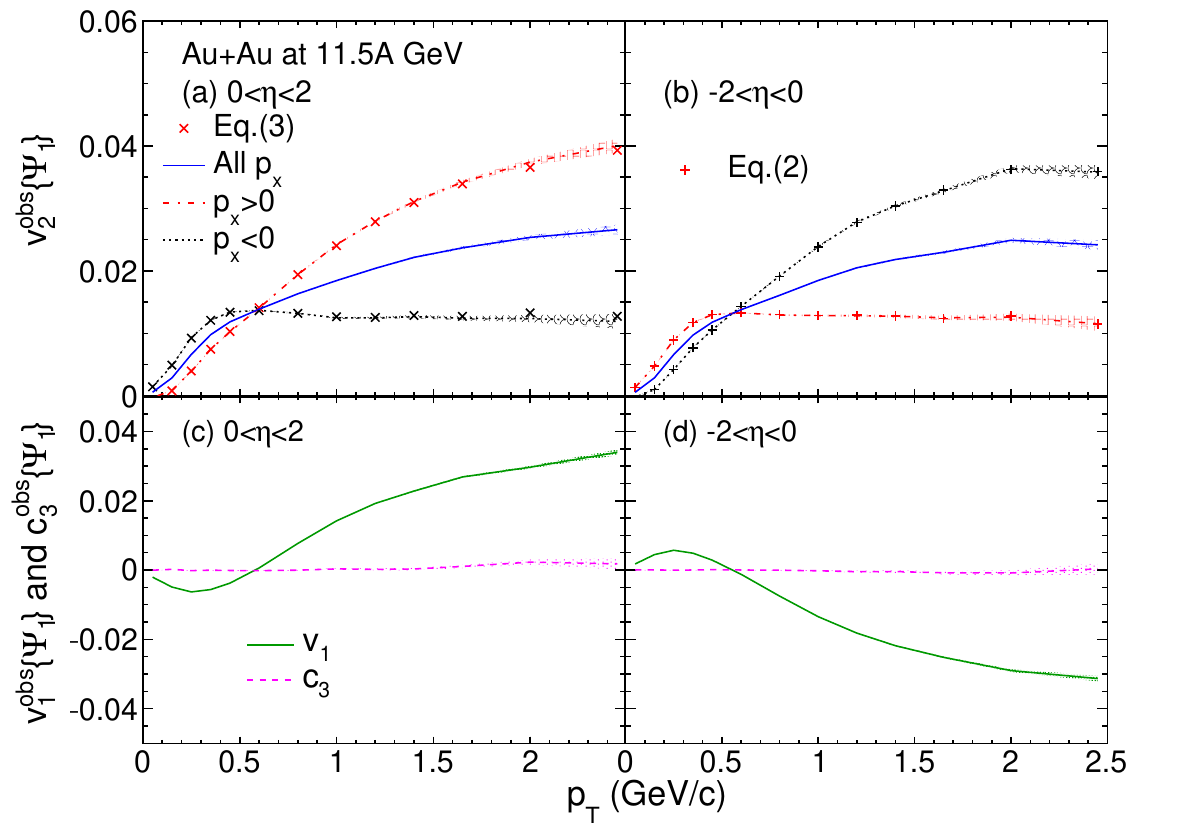}
\caption{Same as Fig.~\ref{fig:5} but without the resolution
  corrections; here the $v_2$ splitting results agree with
  Eqs.\eqref{v2s0}-\eqref{v2s}.} 
\label{fig:6}
\end{figure}

The raw $v_2$ splitting results relative to the first-order event
plane, i.e., without the resolution corrections, are shown in
Fig.~\ref{fig:6}.   We see in panel (a) that the split $v_2$ results
(dot-dashed curve and dotted curve) agree well with the cross symbols,
which are obtained by applying Eq.\eqref{v2s} to the overall $v_1^{\rm
  obs}$ and $v_2^{\rm obs}$ in the left panels. The split $v_2$
results in panel (b) also agree well with the plus symbols that are
obtained by applying Eq.\eqref{v2s0}; this is expected since
$c_3\simeq 0$ in Fig.~\ref{fig:6} and thus Eq.\eqref{v2s0} and
Eq.\eqref{v2s} are essentially the same. 
Therefore, the raw overall $v_1$ extracted with Eq.~\eqref{v2s} from
the split $v_2$ measurements relative to $\Psi_1$ can be corrected 
with its event plane resolution to yield the corrected $v_1$, which
will be equivalent to the standard $v_1\{\Psi_1\}$. It is also
apparent that Eq.\eqref{v2s0} and Eq.\eqref{v2s} fail in
Fig.~\ref{fig:5} because of the different resolution values for $v_1$
and $v_2$. Specifically, $R_1\{\Psi_1\}=0.574, R_2\{\Psi_1\}=0.188$,
and $R_3\{\Psi_1\}=0.047$ are the resolutions used to obtain the
corrected $v_1$, $v_2$, and $c_3$ in Fig.~\ref{fig:5}. In addition, it
is interesting to see the large relative $v_2$ splitting in
Fig.~\ref{fig:6}; this is mostly due to the higher resolution
$R_1\{\Psi_1\}$ than $R_2\{\Psi_1\}$, which makes  $v_1^{\rm
  obs}/v_2^{\rm obs}$ much bigger than the corrected $v_1/v_2$.

\subsection{Relative to the second-order event plane}

Here we explore the $v_2$ splitting analysis relative to the
second-order event plane. First, we need to know the second-order
event plane angle that points toward the direction of the
impact parameter, which we call $\Psi_{2\rm r}$, to distinguish the left
side from the right side. However, the standard $\Psi_2$ does not
contain this  information since $\Psi_2 \in (-\pi/2,\pi/2]$ has a
range of $\pi$ instead of $2\pi$. We thus propose the following
correlator for each event:
\begin{equation}
v_1\{\Psi_2\} = \langle \cos(\phi-\Psi_2)\rangle. 
\end{equation}
In the above, $\phi$ is the azimuthal angle of a particle within the
narrowed EPD $\eta$ range, $\Psi_2$ is the second-order event plane
angle, and the bracket $\langle \rangle$ represents the averaging over
particles in the event. We then use the sign of $v_{1,\rm
  west}\{\Psi_2\}$ (for particles within the narrowed EPD range
$3.5<\eta<5.09$) to determine the directional second-order event plane
angle $\Psi_{2\rm r}$:
\begin{eqnarray}
\begin{split}
\Psi_{2\rm r}&=\Psi_2 {\rm ~~~~~~~~if~} v_{1,\rm west}\{\Psi_2\} \ge 0,\\
&=\Psi_2+\pi {\rm ~~~otherwise.}
\end{split}
\end{eqnarray}
For the $v_2$ splitting analysis, we treat the $\Psi_{2\rm r}$ angle
as the $x$-axis; e.g., any hadron with $\cos(\phi-\Psi_{2\rm r}) > 0$
belongs to the $\pxp$ group. 

\begin{figure}[!htb]
\includegraphics[scale=0.45]{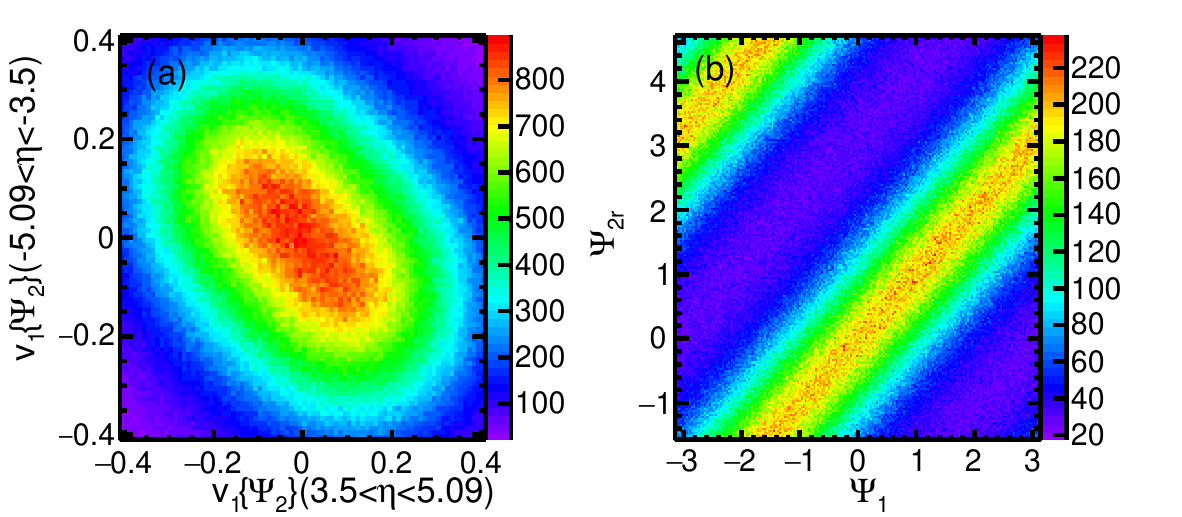}
\caption{AMPT results on the event-by-event (a) $v_1\{\Psi_2\}$ for
  particles within $-5.09<\eta<-3.5$ versus that within
  $3.5<\eta<5.09$, and (b) reconstructed $\Psi_{2\rm r}$ versus
  $\Psi_1$ for minimum bias Au+Au collisions at $11.5A$ GeV.}  
	\label{fig:7}
\end{figure}

Figure~\ref{fig:7}(a) shows the event-by-event correlation between 
$v_1\{\Psi_2\}$ from particles within $3.5<\eta<5.09$ and that from 
particles within $-5.09<\eta<-3.5$ from the AMPT model (with randomized
reaction plane angle) for Au+Au collisions at $11.5A$ GeV.  We see that
$v_{1,\rm west}\{\Psi_2\}$ is significantly anticorrelated with
$v_{1,\rm  east}\{\Psi_2\}$. Figure~\ref{fig:7}(b) shows the
event-by-event correlation between the reconstructed $\Psi_1$ and
$\Psi_{2\rm r}$ angles, which is strong and positive as expected. 
Therefore, we use $\Psi_{2\rm r}$ as the event plane angle for the
$v_2$ splitting analysis in this section, where $v_1$ and $v_2$ are
calculated using the two-subevent method~\cite{Voloshin:2008dg}: 
\begin{equation}
v_n^a\{\Psi_{2\rm r}\} = \frac{ \langle \langle \cos[n(\phi-\Psi_{2\rm r}^b)]
  \rangle \rangle}{R_n^{\rm sub}\{\Psi_{2\rm r}^b\}}.  
\end{equation}
Note that $c_3$ is calculated similarly, and the two-subevent method 
removes the self-correlation between $\Psi_{2\rm r}$ and the flow
coefficients ($v_1$, $v_2$, and $c_3$) calculated relative to
$\Psi_{2\rm r}$. The two subevents here are from $0<\eta<2$ and
$-2<\eta<0$, and the subevent resolutions are calculated as
$R_n^{\rm sub}\{\Psi_{2\rm r}\}=\sqrt {<\cos [n(\Psi_{2\rm
    r}^a-\Psi_{2\rm r}^b)]>}$.

\begin{figure}[!htb]
\includegraphics[scale=0.45]{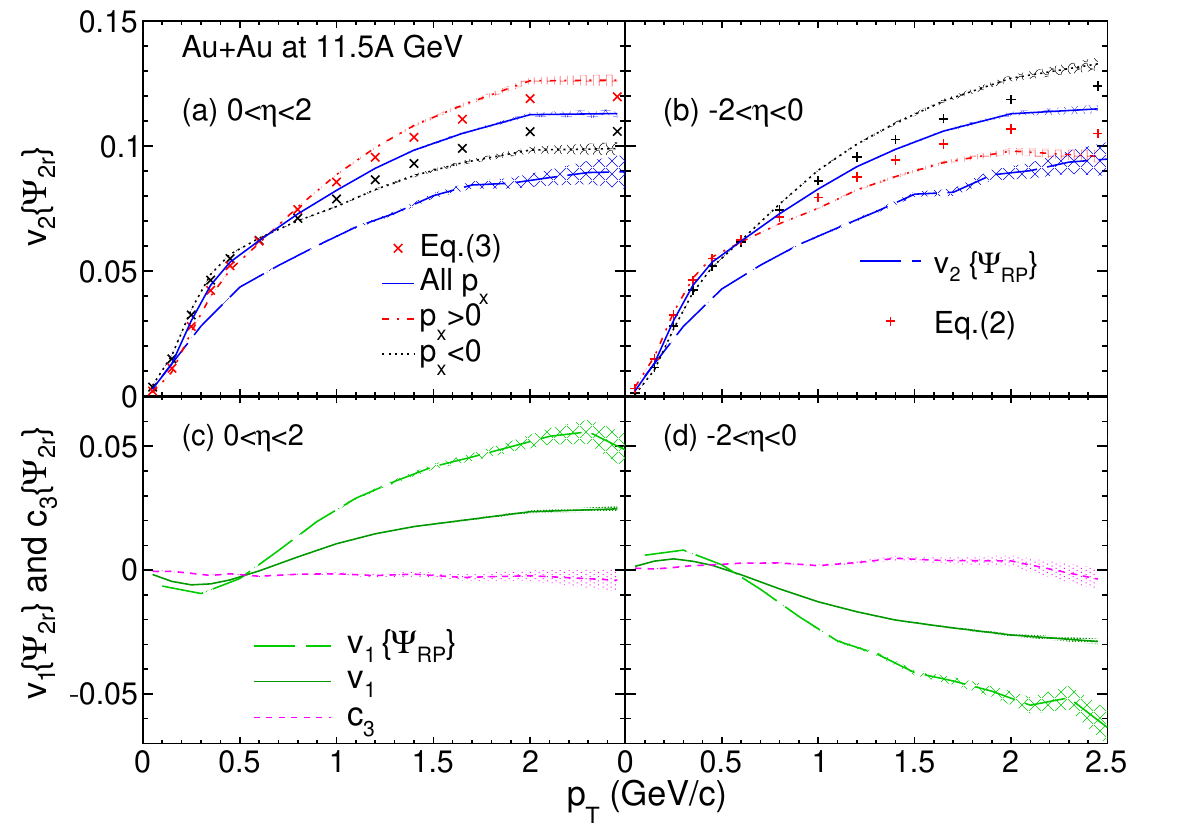}
\caption{Same as Fig.~\ref{fig:5} for results after applying
  resolution corrections but relative to the event plane   angle
  $\Psi_{2\rm r}$.} 
\label{fig:8}
\end{figure}

We show in Fig.~\ref{fig:8}(a) and (b) the $v_2$ splitting results 
of charged hadrons within $\etap$ and $\etan$, respectively, after the
resolution $R_2^{\rm sub}\{\Psi_{2\rm r}\}$ (=0.433 here) is applied. 
The overall $v_1\{\Psi_{2\rm r}\}$ and $c_3\{\Psi_{2\rm r}\}$ results
are shown in Figs.~\ref{fig:8}(c) and (d), after the resolutions
$R_1^{\rm sub}\{\Psi_{2\rm r}\}$ (=0.715 here) and $R_3^{\rm
  sub}\{\Psi_{2\rm r}\}$ (=0.357 here) are applied to correct $v_1$ and 
$c_3$, respectively.  The cross symbols and plus symbols represent
respectively the expectations of Eq.\eqref{v2s} and Eq.\eqref{v2s0}   
based on the corresponding overall corrected $v_2$ and $v_1$. 
Similarly to Fig.~\ref{fig:5}, Eqs.\eqref{v2s0}-\eqref{v2s} fail to
describe the  split $v_2$ curves, and this is due to the different
resolution values. In addition, we see that although $v_2\{\Psi_{2\rm
  r}\}$ are still higher than $v_2\{\Psi_{\rm RP}\}$ (long dashed
curves in the upper panels),  the difference between them is smaller
than that between  $v_2\{\Psi_1\}$ and $v_2\{\Psi_{\rm
  RP}\}$. Furthermore, unlike $v_1\{\Psi_1\}$, $v_1\{\Psi_{2\rm r}\}$
is much lower than $v_1\{\Psi_{\rm RP}\}$  (long dashed curves in the
lower panels).

\begin{figure}[!htb]
\includegraphics[scale=0.45]{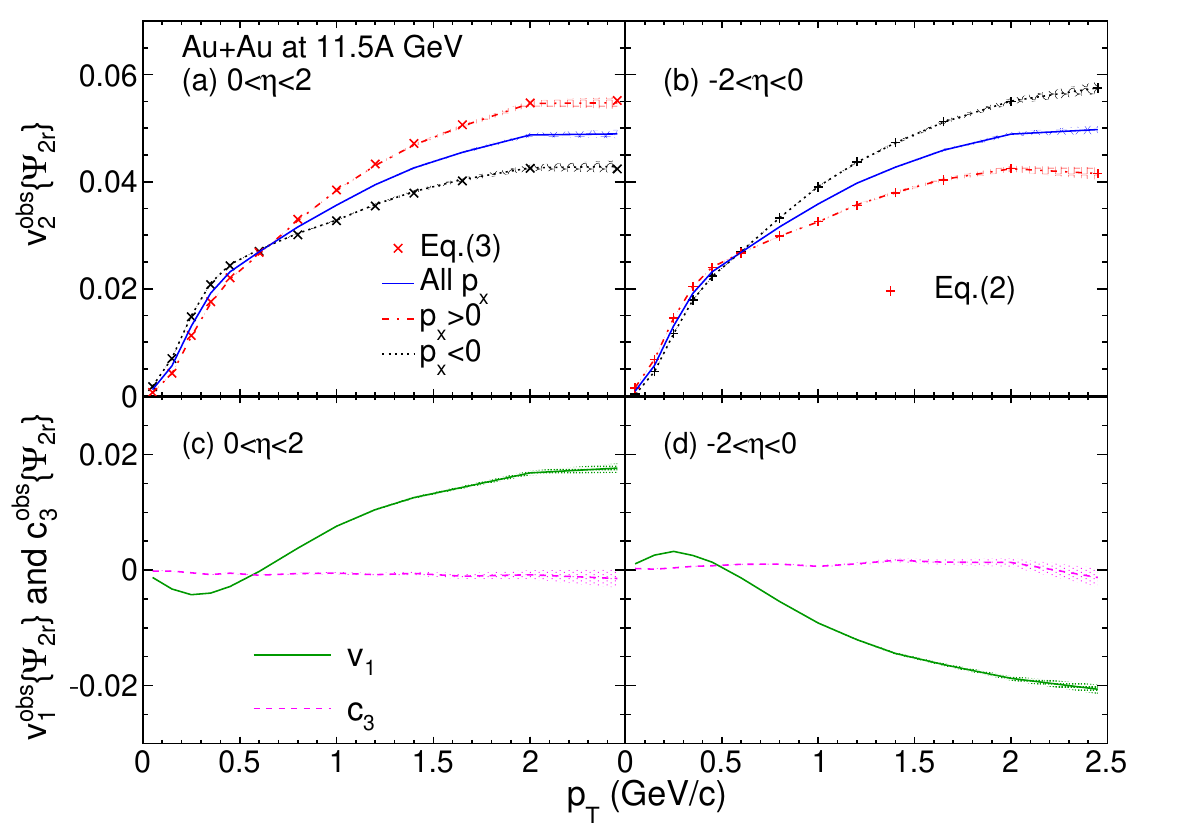}
\caption{Same as Fig.~\ref{fig:6} for results without the resolution
  corrections but relative to the event plane angle $\Psi_{2\rm r}$.} 
\label{fig:9}
\end{figure}

Figure~\ref{fig:9} shows the raw results relative to $\Psi_{2\rm r}$ 
without resolution corrections. 
Similarly to the raw results relative to $\Psi_1$ shown in
Fig.~\ref{fig:6}, the raw $v_2$ split curves relative to $\Psi_{2\rm
  r}$ also agree well with the expectations from
Eqs.\eqref{v2s0}-\eqref{v2s}.  
As a result, the raw overall $v_2$ extracted with Eq.~\eqref{v2s}
from the split $v_2$ measurements relative to $\Psi_{2\rm r}$ can be 
corrected with its event plane resolution to yield the corrected
$v_2$, which will be equivalent to the standard
$v_2\{\Psi_2\}$. Therefore, the $v_2$ splitting measurement should be
feasible  experimentally, using either the first-order event plane
$\Psi_1$ or the directional second-order event plane $\Psi_{2\rm r}$.

\begin{figure}[!htb]
\includegraphics[scale=0.45]{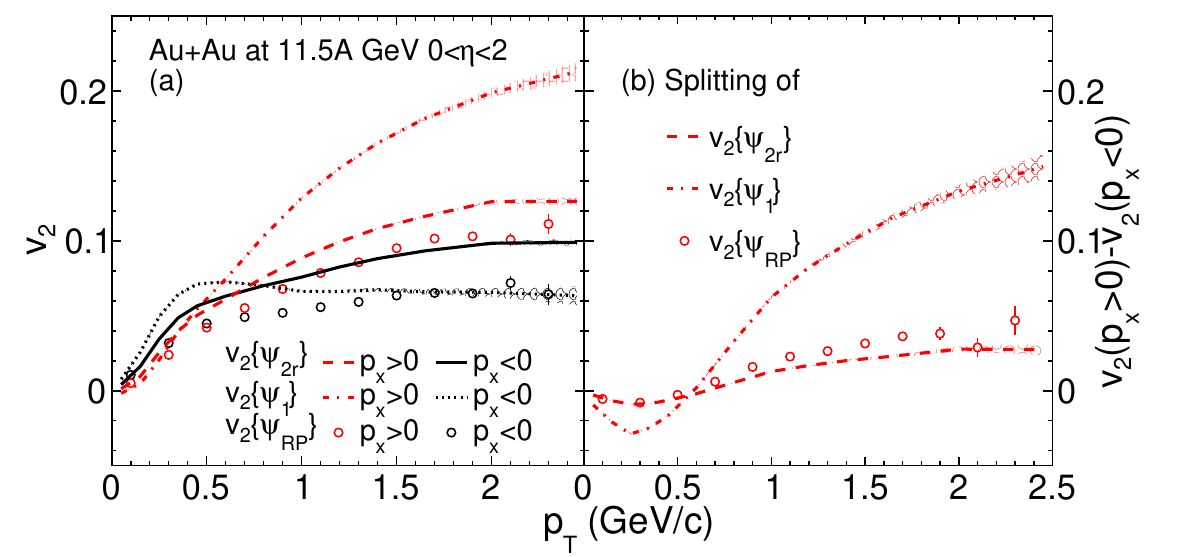}
\caption{(a) The split $v_2$ and (b) the $v_2$ splitting
  $v_2(\pxp)-v_2(\pxn)$ of charged hadrons within $0<\eta<2$ relative 
  to the event plane angle $\Psi_1$, $\Psi_{2\rm r}$, and the theoretical reaction
  plane angle $\Psi_{\rm RP}$ (circles) as functions of the transverse
  momentum  from the  AMPT model for minimum bias Au+Au collisions at
  $11.5A$ GeV.}
\label{fig:10}
\end{figure}

Figure~\ref{fig:10} shows in panel (a) the $v_2(\pxp)$ and $v_2(\pxn)$
results of charged hadrons within $0<\eta<2$ from the AMPT model for
minimum bias Au+Au collisions at $11.5A$ GeV relative to three angles: 
the first-order event plane angle $\Psi_1$, the directional
second-order event plane angle $\Psi_{2\rm r}$, and the theoretical
reaction plane angle  $\Psi_{\rm RP}$. The corresponding results for
the $v_2$ splitting, i.e., $v_2(\pxp)-v_2(\pxn)$, are shown in panel
(b).  We see in panel (a) that the $v_2(\pxp)$ and $v_2(\pxn)$ results
(after applying the resolution correction) relative to the event plane
angle $\Psi_{2\rm r}$ agree better with, although are somewhat higher
than, the results relative to the reaction plane.  From
Fig.~\ref{fig:10} (b), it is clear that the $v_2$ splitting relative
to the event plane angle $\Psi_{2\rm r}$ is much closer to that
relative to the reaction plane. Therefore, the directional
second-order event plane $\Psi_{2\rm r}$ is preferred from the
perspective of measuring the $v_2$ splitting.

\section{Discussions}
\label{Discussions}
 
We emphasize that the string melting AMPT model currently cannot
describe the directed flow well~\cite{Nayak:2019vtn,Nayak:2020djj},
even though the model typically well describes the elliptic and
triangular flows. Therefore, results of the $v_2$ splitting from the
AMPT model in this study are used primarily to demonstrate the
relationship between the $v_2$ splitting and $v_1$ as shown in 
Eqs.\eqref{v2s0}-\eqref{v2s}. The AMPT results of $v_1(\pt)$ and the
$v_2$ splitting here are not intended as quantitative
predictions for the experimental observables; instead, they represent
the exploration of  the $v_2$ splitting possibilities.
The inability of the string melting AMPT model to reproduce 
$v_1$ observables is related to its neglect of the finite (i.e.,
nonzero) nuclear thickness along the beam ($z$) direction, because 
the nuclear $z$-width varies along the $x$-direction and would thus
affect the tilted shape of the created matter.
In addition, the finite nuclear thickness has been shown to
significantly affect the energy density of the dense matter
at low energies~\cite{Lin:2017lcj,Mendenhall:2020fil} and the dynamics
at finite rapidities even at high energies~\cite{Shen:2017bsr}. 

It has been realized in recent years that non-central heavy ion collisions
are affected by other interesting effects such as a strong 
electromagnetic field~\cite{Voronyuk:2011jd,Deng:2012pc}
and vorticity field~\cite{STAR:2017ckg}. 
For example, a significant fraction of the initial total angular
momentum in a non-central heavy ion collision
is deposited into the created dense matter and thus creates a global
vorticity field, which leads to global polarization of hyperons and
spin alignment of vector mesons that have been observed
~\cite{Liang:2004ph,STAR:2017ckg,ALICE:2019aid}. Since we find that
the left-right $v_2$  splitting is mostly due to $v_1$, it will be
interesting to study how the vorticity field affects the development
of $v_1$ and consequently the $v_2$ splitting~\cite{Chen:2021wiv}. 
Note that the global vorticity field and $v_1$ share several 
similarities; e.g., the magnitude is usually larger at larger
rapidities, and they both have a left-right antisymmetry. 

More generally, it will be useful to study the correlations and 
relationships among anisotropic flows, the electromagnetic field, and
the vorticity field. For example, the splitting of charm
and anti-charm directed flows has been proposed as a result of the 
electromagnetic field~\cite{Das:2016cwd}. The rotation of the
created matter has also been shown to affect the flow pattern,
especially the directed flow~\cite{Csernai:2011qq}. 
The dynamics that generates $v_1$ is quite complicated, as $v_1$
depends on the equation of state, mean-field potentials, particle 
rescatterings, shadowing from spectator nucleons, and the tilt of the
created matter in the $x{-}z$ plane
~\cite{Snellings:1999bt,Zhang:2018wlk,Nara:2016phs}. On the other
hand, $v_2$ depends on the geometry  of the created matter in the
$x{-}y$ plane. Therefore, the left-right $v_2$ splitting reflects the
three-dimensional geometry and dynamical evolution of the dense
matter.

\section{Summary}

We have examined the left-right splitting of the elliptic
flow, on opposite sides of the impact parameter axis,
that has been recently proposed as a new effect due to the global
vorticity in non-central heavy ion collisions. 
In this study, we find that this $v_2$ splitting is due to the
non-zero directed flow $v_1$ at finite rapidities and approximately
given by $8v_1(1-3v_2)/(3\pi)$. Therefore, the splitting is expected
to depend sensitively  on the transverse momentum, rapidity range, and
particle species.  

We also use the string melting version of a
multi-phase transport model, which automatically includes the
vorticity field and flow fluctuations and usually describes well the
elliptic and triangular flows. Our model results confirm the existence
of the $v_2$  splitting at finite rapidities on both parton and hadron
levels above certain transverse momenta.  
We also find that the relative splitting may be significant
even for central heavy ion collisions, the splitting within a fixed
(pseudo)rapidity range is expected to decrease with the colliding
energy, and the splitting at large rapidity may be significant even at
LHC energies. In addition, we find that the $v_2$ splitting
should be measurable experimentally with the event plane method 
using either the first- or second-order event plane, 
and that the analytical expectations for the $v_2$ splitting 
apply to the raw $v_2$ and $v_1$ (i.e., before applying their
respective event plane resolutions) in each case. 
Therefore, one can extract the raw overall $v_1$ and $v_2$ 
from the $v_2$ splitting measurements. 
In the typical case where $c_3 \simeq 0$, after applying event plane
resolutions the corrected overall $v_1$ is equivalent to the usual
$v_1$ if the $v_2$ splitting is measured relative to the first-order
event plane,  while the corrected overall $v_2$ is equivalent to the
usual $v_2$ if the $v_2$ splitting is measured relative to the
second-order event plane. 
From the perspective of measuring the split $v_2$, the directional
second-order event plane $\Psi_{2\rm r}$ is preferred since its $v_2$
results agree better with those relative to the reaction plane.  
The left-right $v_2$ splitting, as a complementary observable to the
standard flow observables, will  benefit the studies of the
three-dimensional geometry and dynamical  evolution of the dense
matter created in heavy ion collisions.

\section*{Acknowledgement}
C.Z. acknowledges support from the Chinese Scholarship Council.
This work is supported the National Science Foundation
under Grant No. 2012947 (Z.-W.L.).  

\bibliography{reference}

\providecommand{\noopsort}[1]{}\providecommand{\singleletter}[1]{#1}%
\begin{thebibliography}{46}%
\makeatletter
\providecommand \@ifxundefined [1]{%
 \@ifx{#1\undefined}
}%
\providecommand \@ifnum [1]{%
 \ifnum #1\expandafter \@firstoftwo
 \else \expandafter \@secondoftwo
 \fi
}%
\providecommand \@ifx [1]{%
 \ifx #1\expandafter \@firstoftwo
 \else \expandafter \@secondoftwo
 \fi
}%
\providecommand \natexlab [1]{#1}%
\providecommand \enquote  [1]{``#1''}%
\providecommand \bibnamefont  [1]{#1}%
\providecommand \bibfnamefont [1]{#1}%
\providecommand \citenamefont [1]{#1}%
\providecommand \href@noop [0]{\@secondoftwo}%
\providecommand \href [0]{\begingroup \@sanitize@url \@href}%
\providecommand \@href[1]{\@@startlink{#1}\@@href}%
\providecommand \@@href[1]{\endgroup#1\@@endlink}%
\providecommand \@sanitize@url [0]{\catcode `\\12\catcode `\$12\catcode
  `\&12\catcode `\#12\catcode `\^12\catcode `\_12\catcode `\%12\relax}%
\providecommand \@@startlink[1]{}%
\providecommand \@@endlink[0]{}%
\providecommand \url  [0]{\begingroup\@sanitize@url \@url }%
\providecommand \@url [1]{\endgroup\@href {#1}{\urlprefix }}%
\providecommand \urlprefix  [0]{URL }%
\providecommand \Eprint [0]{\href }%
\providecommand \doibase [0]{http://dx.doi.org/}%
\providecommand \selectlanguage [0]{\@gobble}%
\providecommand \bibinfo  [0]{\@secondoftwo}%
\providecommand \bibfield  [0]{\@secondoftwo}%
\providecommand \translation [1]{[#1]}%
\providecommand \BibitemOpen [0]{}%
\providecommand \bibitemStop [0]{}%
\providecommand \bibitemNoStop [0]{.\EOS\space}%
\providecommand \EOS [0]{\spacefactor3000\relax}%
\providecommand \BibitemShut  [1]{\csname bibitem#1\endcsname}%
\let\auto@bib@innerbib\@empty
\bibitem [{\citenamefont {Shuryak}(1978)}]{Shuryak:1978ij}%
  \BibitemOpen
  \bibfield  {author} {\bibinfo {author} {\bibfnamefont {E.~V.}\ \bibnamefont
  {Shuryak}},\ }\href {\doibase 10.1016/0370-2693(78)90370-2} {\bibfield
  {journal} {\bibinfo  {journal} {Phys. Lett. B}\ }\textbf {\bibinfo {volume}
  {78}},\ \bibinfo {pages} {150} (\bibinfo {year} {1978})}\BibitemShut
  {NoStop}%
\bibitem [{\citenamefont {Voloshin}\ and\ \citenamefont
  {Zhang}(1996)}]{Voloshin:1994mz}%
  \BibitemOpen
  \bibfield  {author} {\bibinfo {author} {\bibfnamefont {S.}~\bibnamefont
  {Voloshin}}\ and\ \bibinfo {author} {\bibfnamefont {Y.}~\bibnamefont
  {Zhang}},\ }\href {\doibase 10.1007/s002880050141} {\bibfield  {journal}
  {\bibinfo  {journal} {Z. Phys. C}\ }\textbf {\bibinfo {volume} {70}},\
  \bibinfo {pages} {665} (\bibinfo {year} {1996})}\BibitemShut {NoStop}%
\bibitem [{\citenamefont {Alver}\ and\ \citenamefont
  {Roland}(2010)}]{Alver:2010gr}%
  \BibitemOpen
  \bibfield  {author} {\bibinfo {author} {\bibfnamefont {B.}~\bibnamefont
  {Alver}}\ and\ \bibinfo {author} {\bibfnamefont {G.}~\bibnamefont {Roland}},\
  }\href {\doibase 10.1103/PhysRevC.82.039903} {\bibfield  {journal} {\bibinfo
  {journal} {Phys. Rev. C}\ }\textbf {\bibinfo {volume} {81}},\ \bibinfo
  {pages} {054905} (\bibinfo {year} {2010})},\ \bibinfo {note} {[erratum: Phys.
  Rev. C 82, 039903 (2010)]}\BibitemShut {NoStop}%
\bibitem [{\citenamefont {Ackermann}\ \emph {et~al.}(2001)\citenamefont
  {Ackermann} \emph {et~al.}}]{STAR:2000ekf}%
  \BibitemOpen
  \bibfield  {author} {\bibinfo {author} {\bibfnamefont {K.~H.}\ \bibnamefont
  {Ackermann}} \emph {et~al.} (\bibinfo {collaboration} {STAR}),\ }\href
  {\doibase 10.1103/PhysRevLett.86.402} {\bibfield  {journal} {\bibinfo
  {journal} {Phys. Rev. Lett.}\ }\textbf {\bibinfo {volume} {86}},\ \bibinfo
  {pages} {402} (\bibinfo {year} {2001})}\BibitemShut {NoStop}%
\bibitem [{\citenamefont {Adler}\ \emph {et~al.}(2003)\citenamefont {Adler}
  \emph {et~al.}}]{PHENIX:2003qra}%
  \BibitemOpen
  \bibfield  {author} {\bibinfo {author} {\bibfnamefont {S.~S.}\ \bibnamefont
  {Adler}} \emph {et~al.} (\bibinfo {collaboration} {PHENIX}),\ }\href
  {\doibase 10.1103/PhysRevLett.91.182301} {\bibfield  {journal} {\bibinfo
  {journal} {Phys. Rev. Lett.}\ }\textbf {\bibinfo {volume} {91}},\ \bibinfo
  {pages} {182301} (\bibinfo {year} {2003})}\BibitemShut {NoStop}%
\bibitem [{\citenamefont {Aamodt}\ \emph {et~al.}(2010)\citenamefont {Aamodt}
  \emph {et~al.}}]{ALICE:2010suc}%
  \BibitemOpen
  \bibfield  {author} {\bibinfo {author} {\bibfnamefont {K.}~\bibnamefont
  {Aamodt}} \emph {et~al.} (\bibinfo {collaboration} {ALICE}),\ }\href
  {\doibase 10.1103/PhysRevLett.105.252302} {\bibfield  {journal} {\bibinfo
  {journal} {Phys. Rev. Lett.}\ }\textbf {\bibinfo {volume} {105}},\ \bibinfo
  {pages} {252302} (\bibinfo {year} {2010})}\BibitemShut {NoStop}%
\bibitem [{\citenamefont {Aidala}\ \emph {et~al.}(2019)\citenamefont {Aidala}
  \emph {et~al.}}]{PHENIX:2018lia}%
  \BibitemOpen
  \bibfield  {author} {\bibinfo {author} {\bibfnamefont {C.}~\bibnamefont
  {Aidala}} \emph {et~al.} (\bibinfo {collaboration} {PHENIX}),\ }\href
  {\doibase 10.1038/s41567-018-0360-0} {\bibfield  {journal} {\bibinfo
  {journal} {Nature Phys.}\ }\textbf {\bibinfo {volume} {15}},\ \bibinfo
  {pages} {214} (\bibinfo {year} {2019})}\BibitemShut {NoStop}%
\bibitem [{\citenamefont {Chatrchyan}\ \emph {et~al.}(2013)\citenamefont
  {Chatrchyan} \emph {et~al.}}]{CMS:2013jlh}%
  \BibitemOpen
  \bibfield  {author} {\bibinfo {author} {\bibfnamefont {S.}~\bibnamefont
  {Chatrchyan}} \emph {et~al.} (\bibinfo {collaboration} {CMS}),\ }\href
  {\doibase 10.1016/j.physletb.2013.06.028} {\bibfield  {journal} {\bibinfo
  {journal} {Phys. Lett. B}\ }\textbf {\bibinfo {volume} {724}},\ \bibinfo
  {pages} {213} (\bibinfo {year} {2013})}\BibitemShut {NoStop}%
\bibitem [{\citenamefont {Dusling}\ \emph {et~al.}(2018)\citenamefont
  {Dusling}, \citenamefont {Mace},\ and\ \citenamefont
  {Venugopalan}}]{Dusling:2017dqg}%
  \BibitemOpen
  \bibfield  {author} {\bibinfo {author} {\bibfnamefont {K.}~\bibnamefont
  {Dusling}}, \bibinfo {author} {\bibfnamefont {M.}~\bibnamefont {Mace}}, \
  and\ \bibinfo {author} {\bibfnamefont {R.}~\bibnamefont {Venugopalan}},\
  }\href {\doibase 10.1103/PhysRevLett.120.042002} {\bibfield  {journal}
  {\bibinfo  {journal} {Phys. Rev. Lett.}\ }\textbf {\bibinfo {volume} {120}},\
  \bibinfo {pages} {042002} (\bibinfo {year} {2018})}\BibitemShut {NoStop}%
\bibitem [{\citenamefont {Mace}\ \emph {et~al.}(2018)\citenamefont {Mace},
  \citenamefont {Skokov}, \citenamefont {Tribedy},\ and\ \citenamefont
  {Venugopalan}}]{Mace:2018vwq}%
  \BibitemOpen
  \bibfield  {author} {\bibinfo {author} {\bibfnamefont {M.}~\bibnamefont
  {Mace}}, \bibinfo {author} {\bibfnamefont {V.~V.}\ \bibnamefont {Skokov}},
  \bibinfo {author} {\bibfnamefont {P.}~\bibnamefont {Tribedy}}, \ and\
  \bibinfo {author} {\bibfnamefont {R.}~\bibnamefont {Venugopalan}},\ }\href
  {\doibase 10.1103/PhysRevLett.121.052301} {\bibfield  {journal} {\bibinfo
  {journal} {Phys. Rev. Lett.}\ }\textbf {\bibinfo {volume} {121}},\ \bibinfo
  {pages} {052301} (\bibinfo {year} {2018})},\ \bibinfo {note} {[erratum: Phys.
  Rev. Lett. 123, 039901 (2019)]}\BibitemShut {NoStop}%
\bibitem [{\citenamefont {He}\ \emph {et~al.}(2016)\citenamefont {He},
  \citenamefont {Edmonds}, \citenamefont {Lin}, \citenamefont {Liu},
  \citenamefont {Molnar},\ and\ \citenamefont {Wang}}]{He:2015hfa}%
  \BibitemOpen
  \bibfield  {author} {\bibinfo {author} {\bibfnamefont {L.}~\bibnamefont
  {He}}, \bibinfo {author} {\bibfnamefont {T.}~\bibnamefont {Edmonds}},
  \bibinfo {author} {\bibfnamefont {Z.-W.}\ \bibnamefont {Lin}}, \bibinfo
  {author} {\bibfnamefont {F.}~\bibnamefont {Liu}}, \bibinfo {author}
  {\bibfnamefont {D.}~\bibnamefont {Molnar}}, \ and\ \bibinfo {author}
  {\bibfnamefont {F.}~\bibnamefont {Wang}},\ }\href {\doibase
  10.1016/j.physletb.2015.12.051} {\bibfield  {journal} {\bibinfo  {journal}
  {Phys. Lett. B}\ }\textbf {\bibinfo {volume} {753}},\ \bibinfo {pages} {506}
  (\bibinfo {year} {2016})}\BibitemShut {NoStop}%
\bibitem [{\citenamefont {Lin}\ \emph {et~al.}(2016)\citenamefont {Lin},
  \citenamefont {He}, \citenamefont {Edmonds}, \citenamefont {Liu},
  \citenamefont {Molnar},\ and\ \citenamefont {Wang}}]{Lin:2015ucn}%
  \BibitemOpen
  \bibfield  {author} {\bibinfo {author} {\bibfnamefont {Z.-W.}\ \bibnamefont
  {Lin}}, \bibinfo {author} {\bibfnamefont {L.}~\bibnamefont {He}}, \bibinfo
  {author} {\bibfnamefont {T.}~\bibnamefont {Edmonds}}, \bibinfo {author}
  {\bibfnamefont {F.}~\bibnamefont {Liu}}, \bibinfo {author} {\bibfnamefont
  {D.}~\bibnamefont {Molnar}}, \ and\ \bibinfo {author} {\bibfnamefont
  {F.}~\bibnamefont {Wang}},\ }\href {\doibase 10.1016/j.nuclphysa.2016.01.017}
  {\bibfield  {journal} {\bibinfo  {journal} {Nucl. Phys. A}\ }\textbf
  {\bibinfo {volume} {956}},\ \bibinfo {pages} {316} (\bibinfo {year}
  {2016})}\BibitemShut {NoStop}%
\bibitem [{\citenamefont {Kurkela}\ \emph {et~al.}(2018)\citenamefont
  {Kurkela}, \citenamefont {Wiedemann},\ and\ \citenamefont
  {Wu}}]{Kurkela:2018ygx}%
  \BibitemOpen
  \bibfield  {author} {\bibinfo {author} {\bibfnamefont {A.}~\bibnamefont
  {Kurkela}}, \bibinfo {author} {\bibfnamefont {U.~A.}\ \bibnamefont
  {Wiedemann}}, \ and\ \bibinfo {author} {\bibfnamefont {B.}~\bibnamefont
  {Wu}},\ }\href {\doibase 10.1016/j.physletb.2018.06.064} {\bibfield
  {journal} {\bibinfo  {journal} {Phys. Lett. B}\ }\textbf {\bibinfo {volume}
  {783}},\ \bibinfo {pages} {274} (\bibinfo {year} {2018})}\BibitemShut
  {NoStop}%
\bibitem [{\citenamefont {Kurkela}\ \emph {et~al.}(2019)\citenamefont
  {Kurkela}, \citenamefont {Wiedemann},\ and\ \citenamefont
  {Wu}}]{Kurkela:2019kip}%
  \BibitemOpen
  \bibfield  {author} {\bibinfo {author} {\bibfnamefont {A.}~\bibnamefont
  {Kurkela}}, \bibinfo {author} {\bibfnamefont {U.~A.}\ \bibnamefont
  {Wiedemann}}, \ and\ \bibinfo {author} {\bibfnamefont {B.}~\bibnamefont
  {Wu}},\ }\href {\doibase 10.1140/epjc/s10052-019-7428-6} {\bibfield
  {journal} {\bibinfo  {journal} {Eur. Phys. J. C}\ }\textbf {\bibinfo {volume}
  {79}},\ \bibinfo {pages} {965} (\bibinfo {year} {2019})}\BibitemShut
  {NoStop}%
\bibitem [{\citenamefont {Weller}\ and\ \citenamefont
  {Romatschke}(2017)}]{Weller:2017tsr}%
  \BibitemOpen
  \bibfield  {author} {\bibinfo {author} {\bibfnamefont {R.~D.}\ \bibnamefont
  {Weller}}\ and\ \bibinfo {author} {\bibfnamefont {P.}~\bibnamefont
  {Romatschke}},\ }\href {\doibase 10.1016/j.physletb.2017.09.077} {\bibfield
  {journal} {\bibinfo  {journal} {Phys. Lett. B}\ }\textbf {\bibinfo {volume}
  {774}},\ \bibinfo {pages} {351} (\bibinfo {year} {2017})}\BibitemShut
  {NoStop}%
\bibitem [{\citenamefont {Heinz}\ and\ \citenamefont
  {Moreland}(2019)}]{Heinz:2019dbd}%
  \BibitemOpen
  \bibfield  {author} {\bibinfo {author} {\bibfnamefont {U.~W.}\ \bibnamefont
  {Heinz}}\ and\ \bibinfo {author} {\bibfnamefont {J.~S.}\ \bibnamefont
  {Moreland}},\ }\href {\doibase 10.1088/1742-6596/1271/1/012018} {\bibfield
  {journal} {\bibinfo  {journal} {J. Phys. Conf. Ser.}\ }\textbf {\bibinfo
  {volume} {1271}},\ \bibinfo {pages} {012018} (\bibinfo {year}
  {2019})}\BibitemShut {NoStop}%
\bibitem [{\citenamefont {Chen}\ \emph {et~al.}(2021)\citenamefont {Chen},
  \citenamefont {Wang}, \citenamefont {Greiner},\ and\ \citenamefont
  {Xu}}]{Chen:2021wiv}%
  \BibitemOpen
  \bibfield  {author} {\bibinfo {author} {\bibfnamefont {Z.}~\bibnamefont
  {Chen}}, \bibinfo {author} {\bibfnamefont {Z.}~\bibnamefont {Wang}}, \bibinfo
  {author} {\bibfnamefont {C.}~\bibnamefont {Greiner}}, \ and\ \bibinfo
  {author} {\bibfnamefont {Z.}~\bibnamefont {Xu}},\ }\href@noop {} {\
  (\bibinfo {year} {2021})},\ \Eprint {http://arxiv.org/abs/2108.12735}
  {arXiv:2108.12735 [hep-ph]} \BibitemShut {NoStop}%
\bibitem [{\citenamefont {Lin}\ \emph {et~al.}(2005)\citenamefont {Lin},
  \citenamefont {Ko}, \citenamefont {Li}, \citenamefont {Zhang},\ and\
  \citenamefont {Pal}}]{Lin:2004en}%
  \BibitemOpen
  \bibfield  {author} {\bibinfo {author} {\bibfnamefont {Z.-W.}\ \bibnamefont
  {Lin}}, \bibinfo {author} {\bibfnamefont {C.~M.}\ \bibnamefont {Ko}},
  \bibinfo {author} {\bibfnamefont {B.-A.}\ \bibnamefont {Li}}, \bibinfo
  {author} {\bibfnamefont {B.}~\bibnamefont {Zhang}}, \ and\ \bibinfo {author}
  {\bibfnamefont {S.}~\bibnamefont {Pal}},\ }\href {\doibase
  10.1103/PhysRevC.72.064901} {\bibfield  {journal} {\bibinfo  {journal} {Phys.
  Rev. C}\ }\textbf {\bibinfo {volume} {72}},\ \bibinfo {pages} {064901}
  (\bibinfo {year} {2005})}\BibitemShut {NoStop}%
\bibitem [{\citenamefont {Zhang}\ \emph {et~al.}(2021)\citenamefont {Zhang},
  \citenamefont {Zheng}, \citenamefont {Shi},\ and\ \citenamefont
  {Lin}}]{Zhang:2021vvp}%
  \BibitemOpen
  \bibfield  {author} {\bibinfo {author} {\bibfnamefont {C.}~\bibnamefont
  {Zhang}}, \bibinfo {author} {\bibfnamefont {L.}~\bibnamefont {Zheng}},
  \bibinfo {author} {\bibfnamefont {S.}~\bibnamefont {Shi}}, \ and\ \bibinfo
  {author} {\bibfnamefont {Z.-W.}\ \bibnamefont {Lin}},\ }\href {\doibase
  10.1103/PhysRevC.104.014908} {\bibfield  {journal} {\bibinfo  {journal}
  {Phys. Rev. C}\ }\textbf {\bibinfo {volume} {104}},\ \bibinfo {pages}
  {014908} (\bibinfo {year} {2021})}\BibitemShut {NoStop}%
\bibitem [{\citenamefont {Lin}\ and\ \citenamefont {Ko}(2002)}]{Lin:2001zk}%
  \BibitemOpen
  \bibfield  {author} {\bibinfo {author} {\bibfnamefont {Z.-W.}\ \bibnamefont
  {Lin}}\ and\ \bibinfo {author} {\bibfnamefont {C.~M.}\ \bibnamefont {Ko}},\
  }\href {\doibase 10.1103/PhysRevC.65.034904} {\bibfield  {journal} {\bibinfo
  {journal} {Phys. Rev. C}\ }\textbf {\bibinfo {volume} {65}},\ \bibinfo
  {pages} {034904} (\bibinfo {year} {2002})}\BibitemShut {NoStop}%
\bibitem [{\citenamefont {Ma}\ and\ \citenamefont {Lin}(2016)}]{Ma:2016fve}%
  \BibitemOpen
  \bibfield  {author} {\bibinfo {author} {\bibfnamefont {G.-L.}\ \bibnamefont
  {Ma}}\ and\ \bibinfo {author} {\bibfnamefont {Z.-W.}\ \bibnamefont {Lin}},\
  }\href {\doibase 10.1103/PhysRevC.93.054911} {\bibfield  {journal} {\bibinfo
  {journal} {Phys. Rev. C}\ }\textbf {\bibinfo {volume} {93}},\ \bibinfo
  {pages} {054911} (\bibinfo {year} {2016})}\BibitemShut {NoStop}%
\bibitem [{\citenamefont {Jiang}\ \emph {et~al.}(2016)\citenamefont {Jiang},
  \citenamefont {Lin},\ and\ \citenamefont {Liao}}]{Jiang:2016woz}%
  \BibitemOpen
  \bibfield  {author} {\bibinfo {author} {\bibfnamefont {Y.}~\bibnamefont
  {Jiang}}, \bibinfo {author} {\bibfnamefont {Z.-W.}\ \bibnamefont {Lin}}, \
  and\ \bibinfo {author} {\bibfnamefont {J.}~\bibnamefont {Liao}},\ }\href
  {\doibase 10.1103/PhysRevC.94.044910} {\bibfield  {journal} {\bibinfo
  {journal} {Phys. Rev. C}\ }\textbf {\bibinfo {volume} {94}},\ \bibinfo
  {pages} {044910} (\bibinfo {year} {2016})},\ \bibinfo {note} {[erratum: Phys.
  Rev. C 95, 049904 (2017)]}\BibitemShut {NoStop}%
\bibitem [{\citenamefont {Li}\ \emph {et~al.}(2017)\citenamefont {Li},
  \citenamefont {Pang}, \citenamefont {Wang},\ and\ \citenamefont
  {Xia}}]{Li:2017slc}%
  \BibitemOpen
  \bibfield  {author} {\bibinfo {author} {\bibfnamefont {H.}~\bibnamefont
  {Li}}, \bibinfo {author} {\bibfnamefont {L.-G.}\ \bibnamefont {Pang}},
  \bibinfo {author} {\bibfnamefont {Q.}~\bibnamefont {Wang}}, \ and\ \bibinfo
  {author} {\bibfnamefont {X.-L.}\ \bibnamefont {Xia}},\ }\href {\doibase
  10.1103/PhysRevC.96.054908} {\bibfield  {journal} {\bibinfo  {journal} {Phys.
  Rev. C}\ }\textbf {\bibinfo {volume} {96}},\ \bibinfo {pages} {054908}
  (\bibinfo {year} {2017})}\BibitemShut {NoStop}%
\bibitem [{\citenamefont {He}\ and\ \citenamefont {Lin}(2017)}]{He:2017tla}%
  \BibitemOpen
  \bibfield  {author} {\bibinfo {author} {\bibfnamefont {Y.}~\bibnamefont
  {He}}\ and\ \bibinfo {author} {\bibfnamefont {Z.-W.}\ \bibnamefont {Lin}},\
  }\href {\doibase 10.1103/PhysRevC.96.014910} {\bibfield  {journal} {\bibinfo
  {journal} {Phys. Rev. C}\ }\textbf {\bibinfo {volume} {96}},\ \bibinfo
  {pages} {014910} (\bibinfo {year} {2017})}\BibitemShut {NoStop}%
\bibitem [{\citenamefont {Zhang}\ \emph {et~al.}(2019)\citenamefont {Zhang},
  \citenamefont {Zheng}, \citenamefont {Liu}, \citenamefont {Shi},\ and\
  \citenamefont {Lin}}]{Zhang:2019utb}%
  \BibitemOpen
  \bibfield  {author} {\bibinfo {author} {\bibfnamefont {C.}~\bibnamefont
  {Zhang}}, \bibinfo {author} {\bibfnamefont {L.}~\bibnamefont {Zheng}},
  \bibinfo {author} {\bibfnamefont {F.}~\bibnamefont {Liu}}, \bibinfo {author}
  {\bibfnamefont {S.}~\bibnamefont {Shi}}, \ and\ \bibinfo {author}
  {\bibfnamefont {Z.-W.}\ \bibnamefont {Lin}},\ }\href {\doibase
  10.1103/PhysRevC.99.064906} {\bibfield  {journal} {\bibinfo  {journal} {Phys.
  Rev. C}\ }\textbf {\bibinfo {volume} {99}},\ \bibinfo {pages} {064906}
  (\bibinfo {year} {2019})}\BibitemShut {NoStop}%
\bibitem [{\citenamefont {Zheng}\ \emph {et~al.}(2020)\citenamefont {Zheng},
  \citenamefont {Zhang}, \citenamefont {Shi},\ and\ \citenamefont
  {Lin}}]{Zheng:2019alz}%
  \BibitemOpen
  \bibfield  {author} {\bibinfo {author} {\bibfnamefont {L.}~\bibnamefont
  {Zheng}}, \bibinfo {author} {\bibfnamefont {C.}~\bibnamefont {Zhang}},
  \bibinfo {author} {\bibfnamefont {S.~S.}\ \bibnamefont {Shi}}, \ and\
  \bibinfo {author} {\bibfnamefont {Z.-W.}\ \bibnamefont {Lin}},\ }\href
  {\doibase 10.1103/PhysRevC.101.034905} {\bibfield  {journal} {\bibinfo
  {journal} {Phys. Rev. C}\ }\textbf {\bibinfo {volume} {101}},\ \bibinfo
  {pages} {034905} (\bibinfo {year} {2020})}\BibitemShut {NoStop}%
\bibitem [{\citenamefont {Xu}\ \emph {et~al.}(2012)\citenamefont {Xu},
  \citenamefont {Chen}, \citenamefont {Ko},\ and\ \citenamefont
  {Lin}}]{Xu:2012gf}%
  \BibitemOpen
  \bibfield  {author} {\bibinfo {author} {\bibfnamefont {J.}~\bibnamefont
  {Xu}}, \bibinfo {author} {\bibfnamefont {L.-W.}\ \bibnamefont {Chen}},
  \bibinfo {author} {\bibfnamefont {C.~M.}\ \bibnamefont {Ko}}, \ and\ \bibinfo
  {author} {\bibfnamefont {Z.-W.}\ \bibnamefont {Lin}},\ }\href {\doibase
  10.1103/PhysRevC.85.041901} {\bibfield  {journal} {\bibinfo  {journal} {Phys.
  Rev. C}\ }\textbf {\bibinfo {volume} {85}},\ \bibinfo {pages} {041901 (R)}
  (\bibinfo {year} {2012})}\BibitemShut {NoStop}%
\bibitem [{\citenamefont {Xu}\ \emph {et~al.}(2014)\citenamefont {Xu},
  \citenamefont {Song}, \citenamefont {Ko},\ and\ \citenamefont
  {Li}}]{Xu:2013sta}%
  \BibitemOpen
  \bibfield  {author} {\bibinfo {author} {\bibfnamefont {J.}~\bibnamefont
  {Xu}}, \bibinfo {author} {\bibfnamefont {T.}~\bibnamefont {Song}}, \bibinfo
  {author} {\bibfnamefont {C.~M.}\ \bibnamefont {Ko}}, \ and\ \bibinfo {author}
  {\bibfnamefont {F.}~\bibnamefont {Li}},\ }\href {\doibase
  10.1103/PhysRevLett.112.012301} {\bibfield  {journal} {\bibinfo  {journal}
  {Phys. Rev. Lett.}\ }\textbf {\bibinfo {volume} {112}},\ \bibinfo {pages}
  {012301} (\bibinfo {year} {2014})}\BibitemShut {NoStop}%
\bibitem [{\citenamefont {Poskanzer}\ and\ \citenamefont
  {Voloshin}(1998)}]{Poskanzer:1998yz}%
  \BibitemOpen
  \bibfield  {author} {\bibinfo {author} {\bibfnamefont {A.~M.}\ \bibnamefont
  {Poskanzer}}\ and\ \bibinfo {author} {\bibfnamefont {S.~A.}\ \bibnamefont
  {Voloshin}},\ }\href {\doibase 10.1103/PhysRevC.58.1671} {\bibfield
  {journal} {\bibinfo  {journal} {Phys. Rev. C}\ }\textbf {\bibinfo {volume}
  {58}},\ \bibinfo {pages} {1671} (\bibinfo {year} {1998})}\BibitemShut
  {NoStop}%
\bibitem [{\citenamefont {Voloshin}\ \emph {et~al.}(2010)\citenamefont
  {Voloshin}, \citenamefont {Poskanzer},\ and\ \citenamefont
  {Snellings}}]{Voloshin:2008dg}%
  \BibitemOpen
  \bibfield  {author} {\bibinfo {author} {\bibfnamefont {S.~A.}\ \bibnamefont
  {Voloshin}}, \bibinfo {author} {\bibfnamefont {A.~M.}\ \bibnamefont
  {Poskanzer}}, \ and\ \bibinfo {author} {\bibfnamefont {R.}~\bibnamefont
  {Snellings}},\ }\href {\doibase 10.1007/978-3-642-01539-7_10} {\bibfield
  {journal} {\bibinfo  {journal} {Landolt-Bornstein}\ }\textbf {\bibinfo
  {volume} {23}},\ \bibinfo {pages} {293} (\bibinfo {year} {2010})}\BibitemShut
  {NoStop}%
\bibitem [{\citenamefont {Adams}\ \emph {et~al.}(2020)\citenamefont {Adams}
  \emph {et~al.}}]{Adams:2019fpo}%
  \BibitemOpen
  \bibfield  {author} {\bibinfo {author} {\bibfnamefont {J.}~\bibnamefont
  {Adams}} \emph {et~al.},\ }\href {\doibase 10.1016/j.nima.2020.163970}
  {\bibfield  {journal} {\bibinfo  {journal} {Nucl. Instrum. Meth. A}\ }\textbf
  {\bibinfo {volume} {968}},\ \bibinfo {pages} {163970} (\bibinfo {year}
  {2020})}\BibitemShut {NoStop}%
\bibitem [{\citenamefont {Nayak}\ \emph {et~al.}(2019)\citenamefont {Nayak},
  \citenamefont {Shi}, \citenamefont {Xu},\ and\ \citenamefont
  {Lin}}]{Nayak:2019vtn}%
  \BibitemOpen
  \bibfield  {author} {\bibinfo {author} {\bibfnamefont {K.}~\bibnamefont
  {Nayak}}, \bibinfo {author} {\bibfnamefont {S.}~\bibnamefont {Shi}}, \bibinfo
  {author} {\bibfnamefont {N.}~\bibnamefont {Xu}}, \ and\ \bibinfo {author}
  {\bibfnamefont {Z.-W.}\ \bibnamefont {Lin}},\ }\href {\doibase
  10.1103/PhysRevC.100.054903} {\bibfield  {journal} {\bibinfo  {journal}
  {Phys. Rev. C}\ }\textbf {\bibinfo {volume} {100}},\ \bibinfo {pages}
  {054903} (\bibinfo {year} {2019})}\BibitemShut {NoStop}%
\bibitem [{\citenamefont {Nayak}(2021)}]{Nayak:2020djj}%
  \BibitemOpen
  \bibfield  {author} {\bibinfo {author} {\bibfnamefont {K.}~\bibnamefont
  {Nayak}} (\bibinfo {collaboration} {STAR}),\ }\href {\doibase
  10.1016/j.nuclphysa.2020.121855} {\bibfield  {journal} {\bibinfo  {journal}
  {Nucl. Phys. A}\ }\textbf {\bibinfo {volume} {1005}},\ \bibinfo {pages}
  {121855} (\bibinfo {year} {2021})}\BibitemShut {NoStop}%
\bibitem [{\citenamefont {Lin}(2018)}]{Lin:2017lcj}%
  \BibitemOpen
  \bibfield  {author} {\bibinfo {author} {\bibfnamefont {Z.-W.}\ \bibnamefont
  {Lin}},\ }\href {\doibase 10.1103/PhysRevC.98.034908} {\bibfield  {journal}
  {\bibinfo  {journal} {Phys. Rev. C}\ }\textbf {\bibinfo {volume} {98}},\
  \bibinfo {pages} {034908} (\bibinfo {year} {2018})}\BibitemShut {NoStop}%
\bibitem [{\citenamefont {Mendenhall}\ and\ \citenamefont
  {Lin}(2021)}]{Mendenhall:2020fil}%
  \BibitemOpen
  \bibfield  {author} {\bibinfo {author} {\bibfnamefont {T.}~\bibnamefont
  {Mendenhall}}\ and\ \bibinfo {author} {\bibfnamefont {Z.-W.}\ \bibnamefont
  {Lin}},\ }\href {\doibase 10.1103/PhysRevC.103.024907} {\bibfield  {journal}
  {\bibinfo  {journal} {Phys. Rev. C}\ }\textbf {\bibinfo {volume} {103}},\
  \bibinfo {pages} {024907} (\bibinfo {year} {2021})}\BibitemShut {NoStop}%
\bibitem [{\citenamefont {Shen}\ and\ \citenamefont
  {Schenke}(2018)}]{Shen:2017bsr}%
  \BibitemOpen
  \bibfield  {author} {\bibinfo {author} {\bibfnamefont {C.}~\bibnamefont
  {Shen}}\ and\ \bibinfo {author} {\bibfnamefont {B.}~\bibnamefont {Schenke}},\
  }\href {\doibase 10.1103/PhysRevC.97.024907} {\bibfield  {journal} {\bibinfo
  {journal} {Phys. Rev. C}\ }\textbf {\bibinfo {volume} {97}},\ \bibinfo
  {pages} {024907} (\bibinfo {year} {2018})}\BibitemShut {NoStop}%
\bibitem [{\citenamefont {Voronyuk}\ \emph {et~al.}(2011)\citenamefont
  {Voronyuk}, \citenamefont {Toneev}, \citenamefont {Cassing}, \citenamefont
  {Bratkovskaya}, \citenamefont {Konchakovski},\ and\ \citenamefont
  {Voloshin}}]{Voronyuk:2011jd}%
  \BibitemOpen
  \bibfield  {author} {\bibinfo {author} {\bibfnamefont {V.}~\bibnamefont
  {Voronyuk}}, \bibinfo {author} {\bibfnamefont {V.~D.}\ \bibnamefont
  {Toneev}}, \bibinfo {author} {\bibfnamefont {W.}~\bibnamefont {Cassing}},
  \bibinfo {author} {\bibfnamefont {E.~L.}\ \bibnamefont {Bratkovskaya}},
  \bibinfo {author} {\bibfnamefont {V.~P.}\ \bibnamefont {Konchakovski}}, \
  and\ \bibinfo {author} {\bibfnamefont {S.~A.}\ \bibnamefont {Voloshin}},\
  }\href {\doibase 10.1103/PhysRevC.83.054911} {\bibfield  {journal} {\bibinfo
  {journal} {Phys. Rev. C}\ }\textbf {\bibinfo {volume} {83}},\ \bibinfo
  {pages} {054911} (\bibinfo {year} {2011})}\BibitemShut {NoStop}%
\bibitem [{\citenamefont {Deng}\ and\ \citenamefont
  {Huang}(2012)}]{Deng:2012pc}%
  \BibitemOpen
  \bibfield  {author} {\bibinfo {author} {\bibfnamefont {W.-T.}\ \bibnamefont
  {Deng}}\ and\ \bibinfo {author} {\bibfnamefont {X.-G.}\ \bibnamefont
  {Huang}},\ }\href {\doibase 10.1103/PhysRevC.85.044907} {\bibfield  {journal}
  {\bibinfo  {journal} {Phys. Rev. C}\ }\textbf {\bibinfo {volume} {85}},\
  \bibinfo {pages} {044907} (\bibinfo {year} {2012})}\BibitemShut {NoStop}%
\bibitem [{\citenamefont {Adamczyk}\ \emph {et~al.}(2017)\citenamefont
  {Adamczyk} \emph {et~al.}}]{STAR:2017ckg}%
  \BibitemOpen
  \bibfield  {author} {\bibinfo {author} {\bibfnamefont {L.}~\bibnamefont
  {Adamczyk}} \emph {et~al.} (\bibinfo {collaboration} {STAR}),\ }\href
  {\doibase 10.1038/nature23004} {\bibfield  {journal} {\bibinfo  {journal}
  {Nature}\ }\textbf {\bibinfo {volume} {548}},\ \bibinfo {pages} {62}
  (\bibinfo {year} {2017})}\BibitemShut {NoStop}%
\bibitem [{\citenamefont {Liang}\ and\ \citenamefont
  {Wang}(2005)}]{Liang:2004ph}%
  \BibitemOpen
  \bibfield  {author} {\bibinfo {author} {\bibfnamefont {Z.-T.}\ \bibnamefont
  {Liang}}\ and\ \bibinfo {author} {\bibfnamefont {X.-N.}\ \bibnamefont
  {Wang}},\ }\href {\doibase 10.1103/PhysRevLett.94.102301} {\bibfield
  {journal} {\bibinfo  {journal} {Phys. Rev. Lett.}\ }\textbf {\bibinfo
  {volume} {94}},\ \bibinfo {pages} {102301} (\bibinfo {year} {2005})},\
  \bibinfo {note} {[erratum: Phys. Rev. Lett. 96, 039901 (2006)]}\BibitemShut
  {NoStop}%
\bibitem [{\citenamefont {Acharya}\ \emph {et~al.}(2020)\citenamefont {Acharya}
  \emph {et~al.}}]{ALICE:2019aid}%
  \BibitemOpen
  \bibfield  {author} {\bibinfo {author} {\bibfnamefont {S.}~\bibnamefont
  {Acharya}} \emph {et~al.} (\bibinfo {collaboration} {ALICE}),\ }\href
  {\doibase 10.1103/PhysRevLett.125.012301} {\bibfield  {journal} {\bibinfo
  {journal} {Phys. Rev. Lett.}\ }\textbf {\bibinfo {volume} {125}},\ \bibinfo
  {pages} {012301} (\bibinfo {year} {2020})}\BibitemShut {NoStop}%
\bibitem [{\citenamefont {Das}\ \emph {et~al.}(2017)\citenamefont {Das},
  \citenamefont {Plumari}, \citenamefont {Chatterjee}, \citenamefont {Alam},
  \citenamefont {Scardina},\ and\ \citenamefont {Greco}}]{Das:2016cwd}%
  \BibitemOpen
  \bibfield  {author} {\bibinfo {author} {\bibfnamefont {S.~K.}\ \bibnamefont
  {Das}}, \bibinfo {author} {\bibfnamefont {S.}~\bibnamefont {Plumari}},
  \bibinfo {author} {\bibfnamefont {S.}~\bibnamefont {Chatterjee}}, \bibinfo
  {author} {\bibfnamefont {J.}~\bibnamefont {Alam}}, \bibinfo {author}
  {\bibfnamefont {F.}~\bibnamefont {Scardina}}, \ and\ \bibinfo {author}
  {\bibfnamefont {V.}~\bibnamefont {Greco}},\ }\href {\doibase
  10.1016/j.physletb.2017.02.046} {\bibfield  {journal} {\bibinfo  {journal}
  {Phys. Lett. B}\ }\textbf {\bibinfo {volume} {768}},\ \bibinfo {pages} {260}
  (\bibinfo {year} {2017})}\BibitemShut {NoStop}%
\bibitem [{\citenamefont {Csernai}\ \emph {et~al.}(2012)\citenamefont
  {Csernai}, \citenamefont {Strottman},\ and\ \citenamefont
  {Anderlik}}]{Csernai:2011qq}%
  \BibitemOpen
  \bibfield  {author} {\bibinfo {author} {\bibfnamefont {L.~P.}\ \bibnamefont
  {Csernai}}, \bibinfo {author} {\bibfnamefont {D.~D.}\ \bibnamefont
  {Strottman}}, \ and\ \bibinfo {author} {\bibfnamefont {C.}~\bibnamefont
  {Anderlik}},\ }\href {\doibase 10.1103/PhysRevC.85.054901} {\bibfield
  {journal} {\bibinfo  {journal} {Phys. Rev. C}\ }\textbf {\bibinfo {volume}
  {85}},\ \bibinfo {pages} {054901} (\bibinfo {year} {2012})}\BibitemShut
  {NoStop}%
\bibitem [{\citenamefont {Snellings}\ \emph {et~al.}(2000)\citenamefont
  {Snellings}, \citenamefont {Sorge}, \citenamefont {Voloshin}, \citenamefont
  {Wang},\ and\ \citenamefont {Xu}}]{Snellings:1999bt}%
  \BibitemOpen
  \bibfield  {author} {\bibinfo {author} {\bibfnamefont {R.~J.~M.}\
  \bibnamefont {Snellings}}, \bibinfo {author} {\bibfnamefont {H.}~\bibnamefont
  {Sorge}}, \bibinfo {author} {\bibfnamefont {S.~A.}\ \bibnamefont {Voloshin}},
  \bibinfo {author} {\bibfnamefont {F.~Q.}\ \bibnamefont {Wang}}, \ and\
  \bibinfo {author} {\bibfnamefont {N.}~\bibnamefont {Xu}},\ }\href {\doibase
  10.1103/PhysRevLett.84.2803} {\bibfield  {journal} {\bibinfo  {journal}
  {Phys. Rev. Lett.}\ }\textbf {\bibinfo {volume} {84}},\ \bibinfo {pages}
  {2803} (\bibinfo {year} {2000})}\BibitemShut {NoStop}%
\bibitem [{\citenamefont {Zhang}\ \emph {et~al.}(2018)\citenamefont {Zhang},
  \citenamefont {Chen}, \citenamefont {Luo}, \citenamefont {Liu},\ and\
  \citenamefont {Nara}}]{Zhang:2018wlk}%
  \BibitemOpen
  \bibfield  {author} {\bibinfo {author} {\bibfnamefont {C.}~\bibnamefont
  {Zhang}}, \bibinfo {author} {\bibfnamefont {J.}~\bibnamefont {Chen}},
  \bibinfo {author} {\bibfnamefont {X.}~\bibnamefont {Luo}}, \bibinfo {author}
  {\bibfnamefont {F.}~\bibnamefont {Liu}}, \ and\ \bibinfo {author}
  {\bibfnamefont {Y.}~\bibnamefont {Nara}},\ }\href {\doibase
  10.1103/PhysRevC.97.064913} {\bibfield  {journal} {\bibinfo  {journal} {Phys.
  Rev. C}\ }\textbf {\bibinfo {volume} {97}},\ \bibinfo {pages} {064913}
  (\bibinfo {year} {2018})}\BibitemShut {NoStop}%
\bibitem [{\citenamefont {Nara}\ \emph {et~al.}(2016)\citenamefont {Nara},
  \citenamefont {Niemi}, \citenamefont {Ohnishi},\ and\ \citenamefont
  {St\"ocker}}]{Nara:2016phs}%
  \BibitemOpen
  \bibfield  {author} {\bibinfo {author} {\bibfnamefont {Y.}~\bibnamefont
  {Nara}}, \bibinfo {author} {\bibfnamefont {H.}~\bibnamefont {Niemi}},
  \bibinfo {author} {\bibfnamefont {A.}~\bibnamefont {Ohnishi}}, \ and\
  \bibinfo {author} {\bibfnamefont {H.}~\bibnamefont {St\"ocker}},\ }\href
  {\doibase 10.1103/PhysRevC.94.034906} {\bibfield  {journal} {\bibinfo
  {journal} {Phys. Rev. C}\ }\textbf {\bibinfo {volume} {94}},\ \bibinfo
  {pages} {034906} (\bibinfo {year} {2016})}\BibitemShut {NoStop}%
\end{thebibliography}%

\end{document}